\documentclass[11pt,a4paper]{article}

\usepackage{epsfig}
\usepackage{graphicx}
\usepackage{epstopdf}
\usepackage[pdf]{pstricks}
\usepackage{amsmath}
\usepackage{amssymb}
\usepackage{verbatim}
\usepackage{bm}
\usepackage{dsfont}
\usepackage[footnotesize]{caption}
\usepackage{subfigure}
\usepackage{cite}
\usepackage{textcomp}
\usepackage{calc}
\usepackage{geometry}
\usepackage{bbm}
\usepackage{xcolor}
\usepackage[utf8]{inputenc}
\usepackage{enumitem}   
\usepackage{setspace}

\usepackage{color}
\definecolor{cred}{RGB}{180,50,40} 
\definecolor{purple}{RGB}{180,90,180} 
\definecolor{darkgreen}{RGB}{0, 100, 0}
\definecolor{commentblue}{RGB}{0, 0, 255}

\usepackage{hyperref}
\hypersetup{colorlinks=true,urlcolor=blue,linkcolor=red,citecolor=green!60!black}

\renewcommand{\baselinestretch}{1.3}

\begin{document}


\noindent May 2019
\hfill DESY 19-089

\vskip 1.5cm

\begin{center}

\bigskip
{\huge\bf
\begin{spacing}{1.1}
The Different Regimes of \\ Axion Gauge Field Inflation
\end{spacing}
}

\vskip 2cm

\renewcommand*{\thefootnote}{\fnsymbol{footnote}}

{\large
Valerie Domcke\,$^{a\,\hspace{-0.25mm}}$
and Stefan Sandner  $^{a,b\,\hspace{-0.25mm}}$
\\[3mm]
{\it{
$^{a}$
Deutsches Elektronen-Synchrotron (DESY), 22607 Hamburg, Germany \\
$^{b}$
II. Institute of Theoretical Physics, University of Hamburg, 22761 Hamburg, Germany
}}}
\end{center}

\vskip 1cm

\renewcommand*{\thefootnote}{\arabic{footnote}}
\newcommand{\axion}{a}
\setcounter{footnote}{0}


\begin{abstract}

In axion gauge field inflation an axion-like particle driving cosmic inflation is coupled to the Chern-Simons density of an Abelian or non-Abelian gauge group.
In the case of a non-Abelian gauge group, this can lead to the formation of a stable, homogeneous and isotropic gauge field background.
We study the dynamics of the inflaton and gauge fields in terms of the two effective coupling parameters: the gauge coupling and the axion decay constant.
Starting from the Bunch-Davies vacuum in the far past, we find that the non-trivial gauge field background arises only significantly \textit{after}
the cosmic microwave background (CMB) scales have left the horizon.
At these scales, the model thus closely resembles Abelian axion inflation, thereby  naturally reconciling the tension of non-Abelian axion gauge field inflation with the latest CMB observations. 
We further consider two exemplary UV-completions of this setup: multiple Peccei-Quinn axions and axion monodromy in string theory.
In both cases we find that the majority of the parameter space is excluded by theoretical or observational constraints. 
The remaining parameter space can be divided into three regimes.
(i) For small gauge couplings we recover natural inflation.
For large gauge couplings the non-Abelian gauge theory either (ii) mimics the Abelian theory or (iii) non-linear interactions prohibit a linear analysis of the gauge field perturbations.

\end{abstract}

\thispagestyle{empty}


\newpage

\setcounter{tocdepth}{1}
{\hypersetup{linkcolor=black}\renewcommand{\baselinestretch}{1}\tableofcontents}


\section{Introduction}

The paradigm of cosmic inflation is stunningly successful in explaining the flatness and homogeneity of our Universe~\cite{Guth:1980zm} as well as the approximately scale invariant inhomogeneities in the cosmic microwave background (CMB)~\cite{Akrami:2018odb}. 
The concrete particle physics model describing the dynamics of inflation is however still very much an open question. A promising candidate for the particle driving inflation (the inflaton) are axion-like particles, whose (approximate) shift-symmetry ensures the required flatness of the scalar potential of inflation. 
 Couplings of the axion-like particle to the Chern-Simons density of a gauge group respect this symmetry, provide a channel to reheat the Universe and moreover can lead to quite remarkable signals, such as a strongly enhanced,  maximally chiral stochastic gravitational wave background (SGWB).

In the context of Abelian gauge groups, the increase of the inflaton velocity during inflation in single field inflation models implies a strong enhancement of the scalar and tensor perturbation spectrum at small scales, whereas the CMB scales (characterized by a small velocity of the inflaton) remain largely unaffected, see e.g.\ Ref.~\cite{Barnaby:2010vf} for an overview. 
This has interesting consequences for the production of primordial black holes~\cite{Linde:2012bt, Domcke:2017fix,Garcia-Bellido:2016dkw} and the search for SGWBs in the frequency band of LIGO, LISA and the Einstein Telescope~\cite{Cook:2011hg,Barnaby:2011qe,Barnaby:2011vw,Anber:2012du,Domcke:2016bkh,Bartolo:2016ami}. 
In this paper our main focus will be on the couplings to non-Abelian (in particular $SU(2)$) gauge fields, coined `chromo-natural inflation' (CNI) in~\cite{Adshead:2012kp}. 
In this case, there can exist a non-trivial homogeneous and isotropic solution for the background gauge field~\cite{Verbin:1989sg,Maleknejad:2011jw,Maleknejad:2011sq,Maleknejad:2012fw,Adshead:2012kp}. 
Depending on if and when this non-trivial solution is realized in the course of inflation, the predictions either closely resemble the Abelian model or reflect the intrinsic non-Abelian nature. 
The latter case in particular allows for the generation of gravitational waves (GWs) at linear order in the gauge field fluctuations, since the presence of a non-vanishing background gauge field implies gauge fields modes with helicity eigenvalue $\pm 2$, which can directly couple to the metric tensor fluctuations~\cite{Dimastrogiovanni:2012st,Dimastrogiovanni:2012ew,Adshead:2013qp,Adshead:2013nka,Domcke:2018rvv}. 
However, the existence of this non-vanishing background field is highly constrained early on in inflation, when the CMB scales exited the horizon, to the point of being excluded by current CMB observations~\cite{Adshead:2013nka} unless particular shapes of the scalar potential are invoked~\cite{Caldwell:2017chz,DallAgata:2018ybl}.

The dynamical emergence of the non-vanishing background solution from a Bunch-Davies vacuum in the infinite past was studied in Ref.~\cite{Domcke:2018rvv}. 
Initially, in the absence of a non-vanishing gauge field background, the model mimics Abelian axion inflation, including an exponential growth of the gauge field fluctuations due to a tachyonic instability driven by the non-zero inflaton velocity.
 Once the fluctuations reach a critical magnitude, they can then dynamically trigger the non-trivial homogeneous and isotropic background solution.
 If this happens only towards the end of of inflation, the CMB scales are well described by a single field inflation model, in accordance with the data, whereas the non-Abelian nature of the model becomes relevant only at small scales. 

In this paper we extend the analysis of Ref.~\cite{Domcke:2018rvv} to cover the entire perturbative parameter space of CNI, which is characterized by two couplings: 
the (perturbative) gauge coupling $g$ associated with the $SU(2)$ gauge group and the effective coupling between the inflaton field and the gauge fields, determined by $g^2/f_a$, where $f_a$ denotes the fundamental axion decay constant. 
We find that in the entire parameter space studied, the non-trivial gauge field background only emerges significantly \textit{after} the CMB scales have exited the horizon, implying that CMB observations cannot distinguish between a coupling to Abelian versus non-Abelian gauge groups. 
This in particular naturally resolves the tension between CNI and the CMB observations. 
On the contrary, direct gravitational wave detectors, such as the Einstein telescope, could distinguish between these two scenarios 
at least in part of the parameter space. 
We illustrate which parts of the parameter space lead to an observable SGWB sourced by 
the Abelian or inherently non-Abelian regime.
Finally, we discuss possible UV completions in two representative settings: by invoking multiple Peccei-Quinn axions or by considering an axion monodromy model as arising in string theory.
 In both cases, we find the regime of strong gauge field backreaction,
sometimes referred to as the {`magnetic drift regime'~\cite{Adshead:2013nka}}, to be incompatible with theoretical constraints.

As all previous analyses of CNI our analysis is based on the linearized equations of motion for the gauge field fluctuations. 
This prohibits a conclusive investigation of the  (phenomenologically possibly most interesting) regime where both effective couplings, $g$ and $g^2/f_a$, are large,
since in this regime the strong growth of the inherently non-Abelian gauge field fluctuations implies the break-down of the perturbative approach. 
We quantify which parts of the parameter space are affected by this limitation (see also Ref.~\cite{Maleknejad:2018nxz} for a related analysis). Combining this with the dynamical emergence of the non-trivial background gauge field solution, we find the regime of intrinsically non-Abelian dynamics under perturbative control to constitute only a small part of the parameter space. 

Envisaging an implementation of CNI with the gauge groups of the Standard Model (SM) of particle physics, we extend these results to a $SU(2) \times U(1)$ gauge theory.
Within the regime of validity of our analysis and for similar values of the two gauge couplings, the gauge field backreaction onto the inflaton dynamics is always dominated by the Abelian gauge group, which does not suffer from the lack of perturbative control within the gauge sector. 
We argue why this result is expected to hold also in the full non-linear theory, which would imply that at least the inflaton dynamics as well as the contributions to the scalar and tensor power spectra sourced by the Abelian fields may be computed without requiring perturbative control in the non-Abelian sector.

The remainder of this paper is organized as follows. We briefly review CNI with a focus on the dynamically emerging gauge field background in Sec.~\ref{sec:axioninflation}. We derive theoretical and phenomenological constraints on the parameter space in Sec.~\ref{sec:constraints}. This includes the limitations of the perturbative analysis of CNI. Section~\ref{sec:results} summarizes the phenomenology and constraints of emerging CNI in the entire parameter space. In Sec.~\ref{sec:su2u1} we give a brief outlook to an implementation of this mechanism in the SM. We conclude in Sec.~\ref{sec:conclusion}. 
The appendix is dedicated to some details on the gauge field background.


\section{Axion Gauge Field Inflation}
\label{sec:axioninflation}

We consider an inflationary stage driven by a pseudo Nambu-Goldstone boson $\axion$. 
The approximate shift-symmetry ensures that the scalar potential $V(\axion)$ associated with this particle is sufficiently flat to guarantee enough time of inflation  to explain the CMB measurements. 
This shift-symmetry in particular is compatible with {the dimension $5$} derivative coupling of $\axion$ to (dark) gauge fields $A_{\mu}$, leading to the following effective Lagrangian,
\begin{align}
\label{eq:axion_inflation_lagrangian}
 \mathcal{L} =  - \sqrt{-|g_{\mu\nu}|} \left( \frac{1}{2}\partial_{\mu}\axion\partial^{\mu}\axion + \frac{1}{4} F_{\mu\nu}F^{\mu\nu} + {\frac{\alpha}{4\pi f_\axion} }\axion F_{\mu\nu}\widetilde{F}^{\mu\nu} + V(\axion)  \right)\,,
\end{align}
The axion - gauge field interaction strength is controlled by the effective coupling $\alpha/f_{a}$ with $\alpha = g^2/(4 \pi)$.
{Note that we work with the units $M_{P}=c=\hbar=1$, where $M_{P}$ refers to the reduced Planck mass.}
The inflaton $\axion$ may be interpreted as an axion-like particle (see section~\ref{sec:constraints} for details), and we will refer to it simply as `axion' in the following. 
For an $SU(2)$ gauge group -- present in chromo-natural inflation~\cite{Adshead:2012kp} -- the electromagnetic field strength tensor is defined as $F_{\mu\nu} \equiv \partial_{\mu}A_{\nu} - \partial_{\nu}A_{\mu} + g\epsilon^{abc}A_{\mu}^{b}A_{\nu}^{c}$. 
The dual is given by $\widetilde{F}^{\mu\nu} \equiv 1/2 \epsilon^{\mu\nu \alpha\beta} F_{\alpha\beta}$. For $\epsilon$ we take the rank-4 Levi-Civita tensor with convention $\epsilon_{0123}\equiv-1/\sqrt{-|g_{\mu\nu}|}$. 
The spacetime is described by the Friedmann-Lema\^itre-Robertson-Walker metric, i.e.\ $\text{d}s^2 = -\text{d}t^2 + R(t)^2\text{d}\mathbf{x}^2 = R(\tau)^2[-\text{d}\tau^2 + \text{d}\mathbf{x}^2]$ in physical and conformal time respectively.
The cosmological scale factor is denoted by $R$ and we use the convention $R=1$ at the end of inflation.

To proceed, we decompose the axion as well as the gauge fields around an homogeneous and isotropic background as~\cite{Maleknejad:2011jw,Maleknejad:2011sq, Adshead:2012kp, Domcke:2018rvv }
\begin{align}
\label{eq:decomposephi}
    \axion(t,\mathbf{x}) &=  \axion(t)  + \delta \axion(t,\mathbf{x})\,,\\
\label{eq:decomposegamma}
    A^{b}_{i}(t,\mathbf{x}) &= R(t)\psi(t)\delta_{i}^{b} + \delta A_{i}^{b}(t,\mathbf{x})\,,
\end{align}
where $b$ refers to the gauge indices and $i$ to the spatial indices, both taking the values $(1,2,3)$. 
With this ansatz, we obtain the homogeneous equations of motion (EOMs) 
\begin{align}
\label{eq:su2fulleom}
     \axion'' - 3\axion' \left( 1 - \frac{H'}{3H} \right) + \frac{\partial_{\axion}V(\axion)}{H^2} &=  {- \frac{3g\alpha}{\pi f_\axion}\psi^2\left(\frac{\psi}{H} - \frac{\psi'}{H} \right)\,, }\\
\label{eq:eomgamma}
     \psi'' - 3  \psi' \left(1 - \frac{H'}{3H} \right) + \psi \left(2 - \frac{H'}{H} \right) + 2g^2\frac{\psi^3}{H^2} &=  {\frac{g\alpha}{\pi f_\axion} \psi^2 \frac{\axion'}{H}\,.}
\end{align}
Here and in the rest of the paper, we refer with $(')$ to the derivative with respect to the number of e-folds $\text{d}N_{e} \equiv -H\text{d}t$.
Also, we choose the convention $\axion < 0, \axion' < 0$.
The Hubble parameter $H \equiv \left( \frac{\text{d}}{\text{d}t}\ln R \right)$ is fixed by the $00$ component of the Einstein equation
\begin{align}
\label{Hubble}
    3H^2 =  \frac{3}{2}H^2 \left( \psi - \psi' \right)^2 + \frac{3}{2}g^2\psi^4 + H^2\frac{(\axion')^2}{2} + V(\axion)\,.
\end{align}

In the slow roll limit Eq.~(\ref{eq:eomgamma}) may be solved analytically as first proven in~\cite{Adshead:2012kp}.
We demonstrate in appendix~\ref{app:approximation} that the derived formula is indeed a good approximation for the full coupled dynamics of Eqs.~(\ref{eq:su2fulleom}),~(\ref{eq:eomgamma}) in the context of emerging chromo-natural inflation\footnote{The analytical evaluation of Eq.~(\ref{eq:eomgamma}) in \cite{Adshead:2012kp} relies on the interpretation of an initially present non-zero vacuum expectation value for the gauge field $\psi$. 
Here on the contrary we consider the background evolving dynamically from the Bunch-Davies vacuum and show for this case that the slow-roll solution of Eq.~(\ref{eq:eomgamma}) resembles the numerical result to good accuracy.}~\cite{Domcke:2018rvv}. 
We will now explain the concept which is behind this notion.
Therefore, let us summarize the detailed derivation of the analytic approximation from~\cite{Domcke:2018rvv}.
We anticipate that the solutions can be qualitatively divided into two distinct forms:
The function $\psi$ (i) approaches zero or (ii) approaches a positive constant.
The explicit form at late times in de Sitter spacetime -- where the relation $\tau = -1/(RH)$ holds -- is given by
\begin{align}
\label{eq:gammasol}
    \psi = H\frac{c_{i}\xi}{g}\,,
\end{align}
with the three different solutions
\begin{align}
\label{eq:csolutions}
    c_{0} = 0\,, ~~~~~~ c_{1} = \frac{1}{2}\left(1 - \sqrt{1- \left(2 /\xi \right)^2 } \right)\,, ~~~~~~ c_{2} = \frac{1}{2}\left(1 + \sqrt{1- \left(2 /\xi \right)^2 } \right)\,.
\end{align}
In these equations we introduced the dimensionless $\xi$ parameter, which basically encodes the axion velocity but its real significance will become clear soon, defined by
\begin{align}
\label{eq:xiparameter}
\xi \equiv {\frac{\alpha |\axion'|}{2\pi f_\axion}\,.}
\end{align}
The $c_{1,2}$ solutions can only emerge for $\xi \geq 2$ (reflecting that the background field is real valued), and then we have the ordering $c_{0} < c_{1} < c_{2}$.

It is now natural to ask how the background evolves over the course of inflation, where $\xi$ typically increases, {but is bounded from above at CMB scales~\cite{Barnaby:2011qe}}.
In particular, we are interested for which initial conditions the gauge field may develop a stable non-zero vacuum expectation value (vev). We derive in the following the approximate time when the gauge field reaches such a configuration in a {semi-}classical approach.
In a complete picture the background field would dynamically evolve in the presence of the {quantum} fluctuations. 
However, as we are here working in the homogeneous field limit we can only approximate this behaviour. 
So, we neglect these quantum fluctuations during the derivation and include them argumentatively in the final result.
Ref.~\cite{Domcke:2018rvv} showed that the background field undergoes an oscillatory phase at the beginning of inflation\footnote{We note, that this is not true in the presence of fluctuations, since the growing super-horizon fluctuations make the classical approach invalid. However, as we are anyway more interested in the last stage of inflation and the influence of the initial condition on this behaviour, we refrain from a detailed discussion of the far past.}.
Thus, the task is to estimate when the oscillatory phase ends and the background approaches the late time solution given in Eq.~(\ref{eq:gammasol}).
One can show that the solution at early times is characterized by two constants, an amplitude $\omega \geq 0$ and a phase $u_{0} \in [0,4K(-1))$, where $K(-1) \simeq 1.3$ denotes the complete elliptical integral of the first kind.
They are uniquely determined once the initial conditions are fixed. 
In the infinite past the background field can be well approximated with the Jacobian elliptic function to be $\psi \sim \omega/R ~ sn(u_{0}-\omega g/(RH),-1)$,
whereas for late times the gauge field is well approximated by Eq.~(\ref{eq:gammasol}), see Ref.~\cite{Domcke:2018rvv} for details.
The 
oscillatory regime then is simply obtained by comparing both expressions and corresponds to $c_{i} \xi H/g \ll \omega/R$.
This leads to a suitable criterion for which times the background field strongly oscillates in the case of $\omega > 0$
\begin{align}
    \label{eq:omegaparameter}
    - \tau\omega g\gg c_{i} \xi\,.
\end{align}

\begin{figure}[!h]
    \centering
    \subfigure{\includegraphics[width = 7cm]{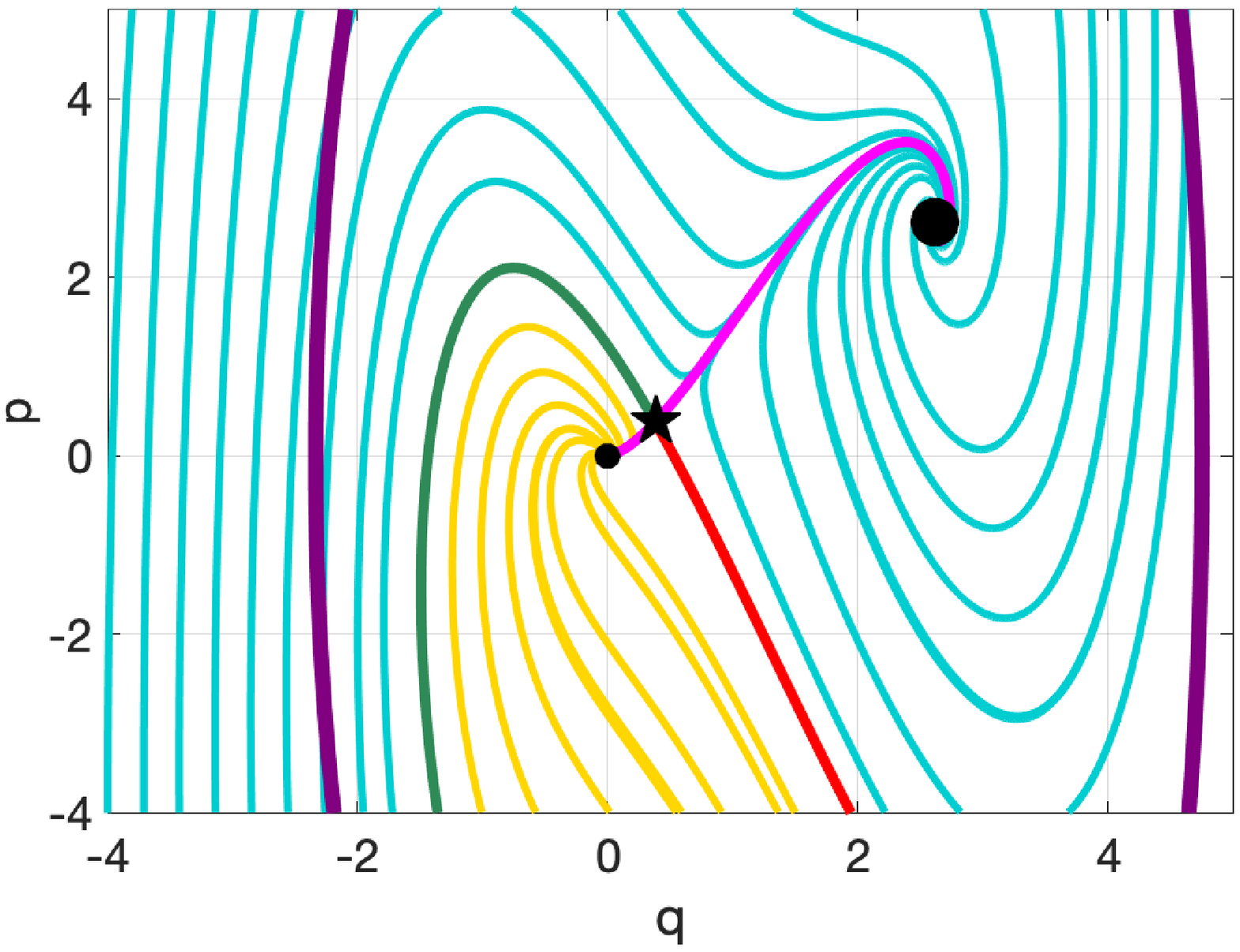}}
    \subfigure{\includegraphics[width = 7cm]{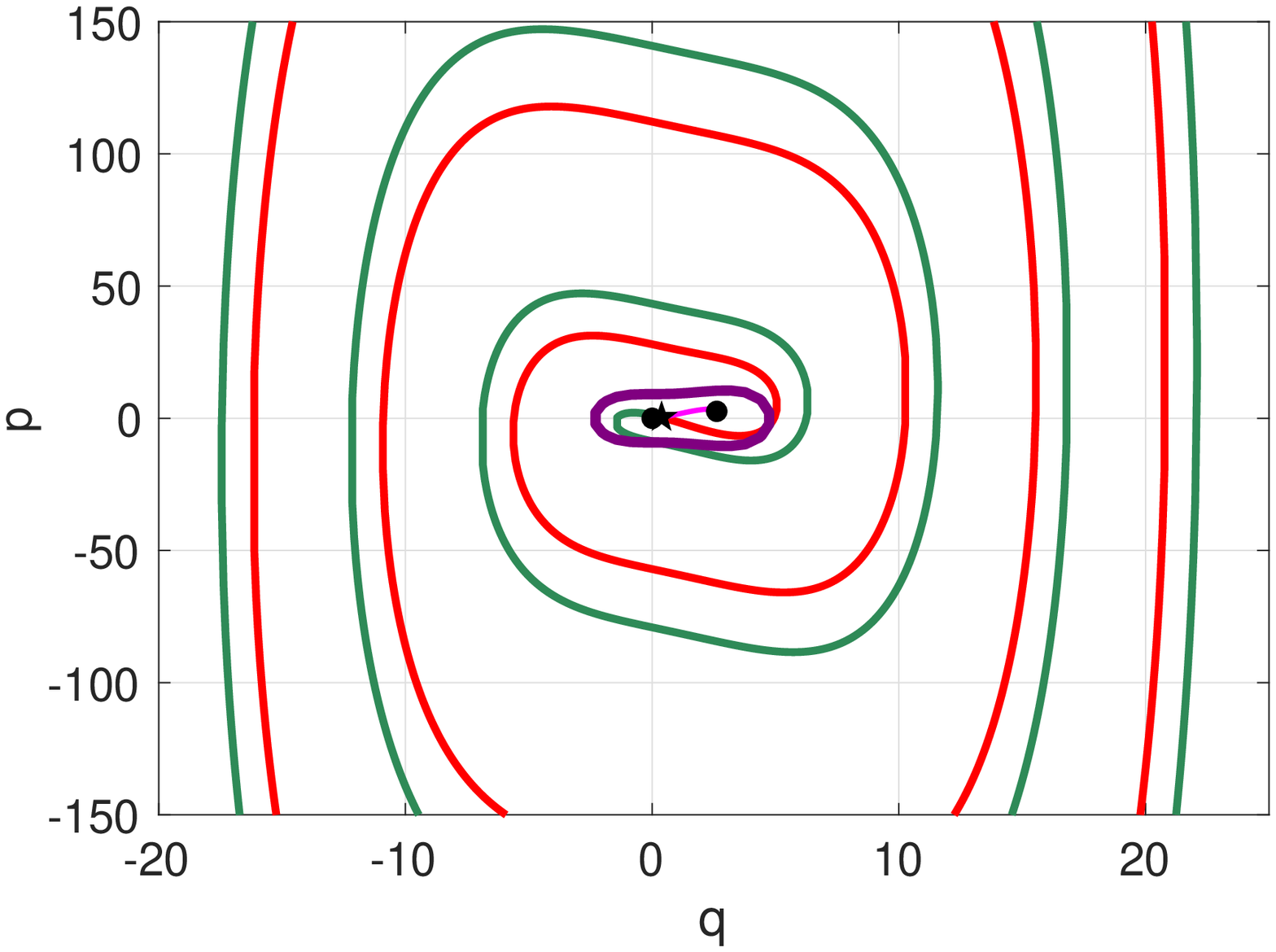}}
    \caption{ Flow of the vector field of the autonomous gauge field background differential equation for $\xi = 3$.  The zeros of the field are given by $\mathbf{z}_{0} = 0$ (small dot), $\mathbf{z}_{1} = c_{1} (\xi,\xi)$ (star) and $\mathbf{z}_{2} = c_{2} (\xi,\xi)$ (big dot). Yellow and blue flow lines correspond to an evolution into the $\mathbf{z}_{1}$ and $\mathbf{z}_{2}$ zero respectively. The magenta line indicates a possible tunneling between the different zeros. Further, we especially highlight with red and green the one parameter family of flow lines evolving into the unstable $\mathbf{z}_{1}$ zero. These two lines mark the boundary for which a randomly located point evolves into either one of the $\mathbf{z}_{0,2}$ zeros. For reference, the purple line denotes the transition regime $\mathbf{-} $ after transforming into the new coordinates $\mathbf{-} $ between an oscillatory and constant solution. This transition happens within $\sim 2$ e-folds once this line is reached.}
    \label{fig:flowlines}
\end{figure}

Seeking a better intuitive understanding what the transition regime actually means, we may bring the background field EOM into an autonomous form\footnote{Autonomous means that the independent variable does not explicitly appear in the differential equation. }.
This gives us the possibility to visualize the dynamics in a simple $2d$ phase space. 
We can choose  the transformation $(q,p) \equiv (-gR\psi \text{e}^{N_{e}},g (R\psi)' \text{e}^{N_{e}})$ which brings it to the desired form, i.e.
\begin{align}
    \frac{\text{d}q}{\text{d}N_{e}} = q - p\,, ~~~~~ \frac{\text{d}p}{\text{d}N_{e}} = 2(q^3-\xi q^2 + p)\,.
\end{align}
To obtain this coupled system we used the EOM for $\psi$ in its de Sitter spacetime form $(R\psi)'' = (R\psi)' - 2\exp(2N)g^2R^3\psi^3 - 2\xi gR^2\psi^2\exp(N)$, c.f. Eq.~(\ref{eq:eomgamma}).
The solution can be visualized by displaying the flow of the vector field $(q-p,2(q^3-\xi q^2 + p))$, see figure~\ref{fig:flowlines}.
The zeros of the vector field are given by $\mathbf{z}_{i} = c_{i}(\xi,\xi)$. 
At early times the lines circuit the zeros of the vector field, marking the oscillatory regime.
Once they cross the border regime of $-\tau\omega g \sim c_{2} \xi$,  they very quickly approach one of the $\mathbf{z}_{i}$ attractors\footnote{{The transition regime can be approximated with the envelope function $T= p^2 + q^4 - (4/3)\xi q^3$ in the $(p,q)$ plane, see Ref.~\cite{Domcke:2018rvv} for details.}}.
We especially highlight the two allowed $c_{1}$ flow lines\footnote{By (a particular $c_{i}$) flow line we mean the line of the vector field 
connected to
one of the $\mathbf{z}_{i}$ zeros of the vector field. No two of such flow lines are allowed to cross, making any flow line path unique.}.
There are only exactly two such lines, revealing that the $c_{1}$ type solution forms a one-parameter family, thus being unstable under perturbations. 
This is contrary to the $c_{0,2}$ type solutions which form a two-parameter family, giving rise to an infinite set of stable trajectories. 
The phase space study reveals a neat structure of the trajectories. 
In particular, the two $c_{1}$ flow lines form the border for which trajectories evolve into either the $\mathbf{z}_{0}$ or $\mathbf{z}_{2}$ {zero}.
It is easily seen that the region of $\mathbf{z}_{2}$ trajectories occupy the majority of the phase space, hence they are rather stable under perturbations.   
So, a randomly placed point in the $(p,q)$ plane inside the oscillatory regime evolves to high probability into the $\textbf{z}_{2}$ {zero} by following the $c_{2}$ flow lines. We note that the probability is dependent on the only free parameter $\xi$ and it increases with increasing $\xi$.
In fact, we show in appendix~\ref{app:approximation}  that the transition from the oscillatory to the constant behaviour of the gauge field background happens in less than $\sim 2$ e-folds when the condition~(\ref{eq:omegaparameter}) is reached.

With this understanding, let us now include the background field fluctuations which we have neglected so far.
{Fluctuations act like perturbations for the field moving along a certain flow line.}
However, so far we kept the discussion general without specifying the initial conditions.
In order to connect to a physical plausible scenario we need to define the initial conditions from which the gauge field starts its evolution.
Therefore, we make use of the property that in de Sitter spacetime all modes $k$ have time-independent frequencies in the infinite past.
The resulting physical unique vacuum is of Bunch-Davies form~\cite{Baumann:2009ds}
\begin{align}
\lim_{\tau\to-\infty} A(k,\tau) = \frac{\text{e}^{-ik\tau}}{\sqrt{2k}}\,,
\end{align}
which is equivalent to the absence of a gauge field background.
Hence, we may consider the gauge field to be initially sitting in the $\textbf{z}_{0}$ zero, c.f. Fig.~\ref{fig:flowlines}. This may be interpreted as the gauge field being initially placed in a local minimum of a potential.
It is stable under small perturbations, but when the threshold of $-\tau \omega g \sim \xi c_{2}$ is reached, the background field {quickly} 'falls' into the $\mathbf{z}_{2}$ zero {(global minimum)} and is thus well described by the $c_{2}$ solution.
In a more complete picture, we would expect that the transition is sourced by the growing quantum fluctuations of the background field.
It is thus natural to replace in first approximation the classical amplitude $\omega$ with its quantum analogue, leading to the transition time
\begin{align}
    \label{eq:transition}
    -\tau g\sqrt{\langle A^2} \rangle \sim c_{2} \xi\,.
\end{align}
We see that the gauge field background can naturally develop a stable non-zero vev, even when starting from the Bunch-Davies vacuum, if the quantum fluctuations $\langle A^2 \rangle^{1/2}$ grow faster than $1/(- \tau)$.
 No additional mechanism or particular initial condition is required, 
contrary to earlier CNI works, see e.g.~\cite{Dimastrogiovanni:2012st,Dimastrogiovanni:2012ew,Adshead:2012kp}. 

Having shown how gauge field fluctuations may source the non-vanishing gauge field background of CNI,
 let us now turn to them in more detail.
To this end, we express the Fourier modes of the non-Abelian fluctuations in terms of the helicity basis $\hat{g}_{\lambda}$
\begin{align}
\delta A^{b}_{\mu}(k,\tau) = \sum_{\lambda} (\hat{g}_{\lambda})^{b}_{\mu} \frac{\omega_{\lambda}({k, \tau})}{\sqrt{2k}}\,,
\end{align}
{where $\lambda$ denotes all six helicity states.}
Particularly interesting for the dynamics are growing modes, {as they dominantly contribute to Eq.~(\ref{eq:transition}). }
{Only the $+2$ helicity mode exhibits a tachyonic instability leading to an exponential growth.}
All the other modes are not enhanced and are thus sub-dominant.
The linearized EOM for this helicity mode is given by~\cite{Domcke:2018rvv}
\begin{align}
\label{eq:ode+2mode}
    \frac{d^2}{dx^2}\omega^k_{+2}(x) + \left( 1-\frac{2\xi}{x} + 2\left( \frac{\xi}{x}-1\right)  \frac{g\psi}{xH} \right)\omega^k_{+2}(x) = 0\,,
\end{align}
where we have introduced $x = - k \tau$. 
Here, the comoving wave number $k$ of the fluctuation exiting the horizon at e-fold $N_{e}$ is given by $k(N_{e}) = H(N_{e}) \text{e}^{-N_{e}}$.
This implies $x = 1$ for the time of horizon exit, and $x <1$ for super-horizon modes.
{Since $\xi$, $\psi$ and $H$ (slowly) evolve during inflation, Eq.~\eqref{eq:ode+2mode} has to be evaluated for a range of modes exiting the horizon at different e-folds.}

Notably, for $c_{i} = c_{0} = 0$ this equation coincides with the one known from the Abelian case~\cite{Anber:2009ua}.
The mode is tachyonically enhanced with the negative effective mass squared term $m^2 = 1-2\xi/x$, purely controlled by the $\xi$ parameter.
Hence, we will refer to this stage as the 'Abelian limit'.
Note that the EOM~(\ref{eq:ode+2mode}) has the general form of the confluent hypergeometric ordinary differential equation.
It may be solved analytically in the slow-roll limit of constant $\xi$ to
\begin{align}
\label{eq:abelmode}
\omega_{\text{AB}}(x) = \text{e}^{\pi\xi/2} W_{-i\xi,1/2}(-2ix),
\end{align}
where $W_{l,m}(n) \equiv \text{e}^{-n/2}n^{m+1/2} U(m-l+1/2,1+2m,n)$ denotes the Whittaker function in terms of the confluent hypergeometric function of the second kind.
It was shown in Ref.~\cite{Cuissa:2018oiw} that this closely resembles also the $\xi \neq \text{const.}$ dynamics in the slow roll regime.
So, the fluctuation required to determine the background field transition time in Eq.~(\ref{eq:transition}) can be well approximated in the Abelian limit as
\begin{align}
 \sqrt{\langle A^{2}_{\text{AB}} \rangle} = \frac{RH}{2\pi} \left(  \int\text{d}x ~x\omega_{\text{AB}}^2\right)^{1/2} \sim \frac{8\times 10^{-3}}{-\tau}\text{e}^{2.8\xi}\,.
\label{eq:variance_ab}
 \end{align}
The integration limits have to be chosen carefully to avoid counting the infinite vacuum energy. 
Most of the integral contribution arises from the region $1/(8\xi) \leq x \leq 2\xi$, such that this will give us a good approximation to the full solution.
{We note that for monotonously increasing $\xi$, the fluctuations grow faster than $1/(-\tau)$, as required to dynamically source a non-trivial gauge field background.}

\begin{figure}[!h]
    \centering
    \subfigure{\includegraphics[width = 7cm]{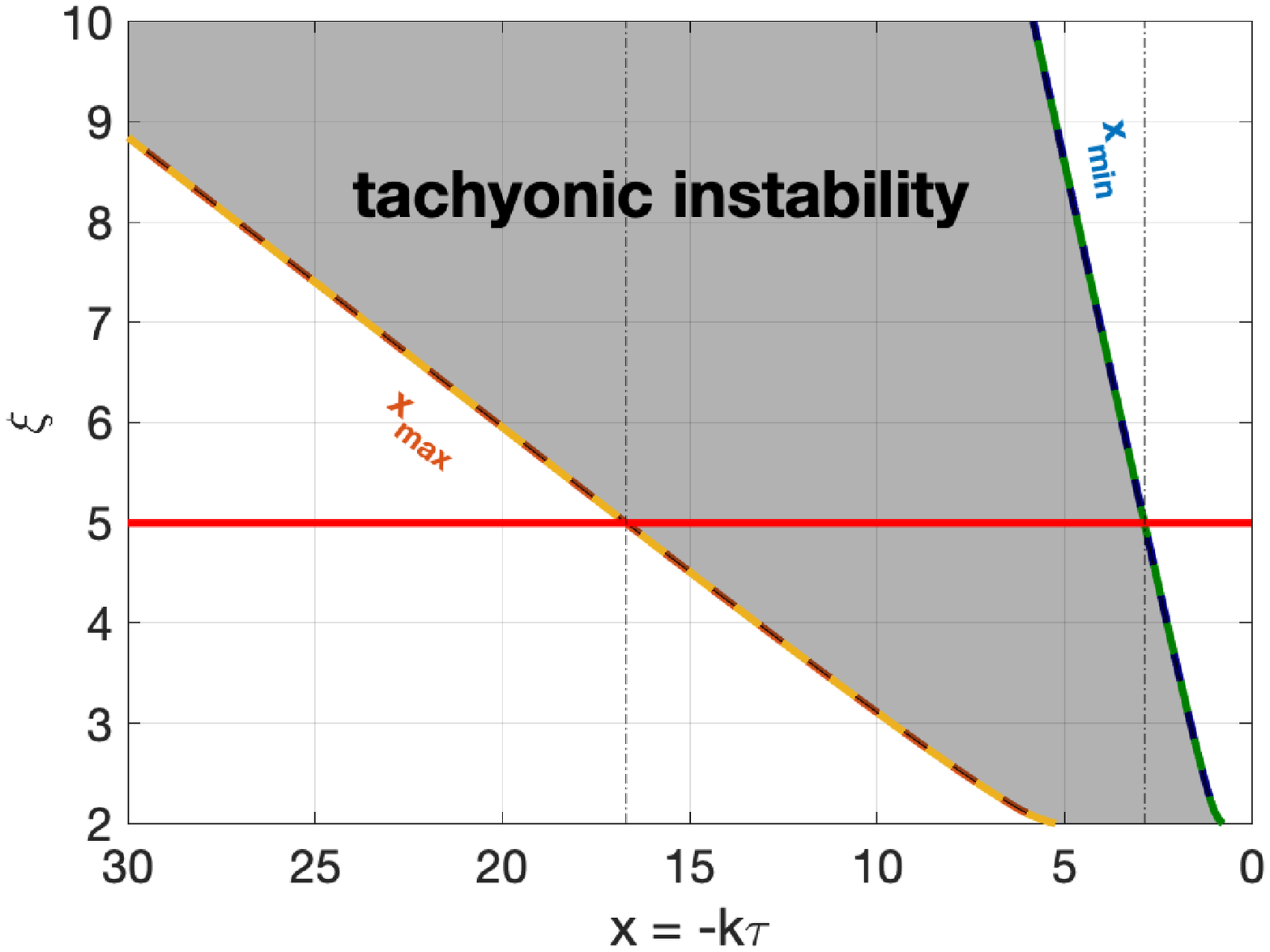}}
    \subfigure{\includegraphics[width = 7cm]{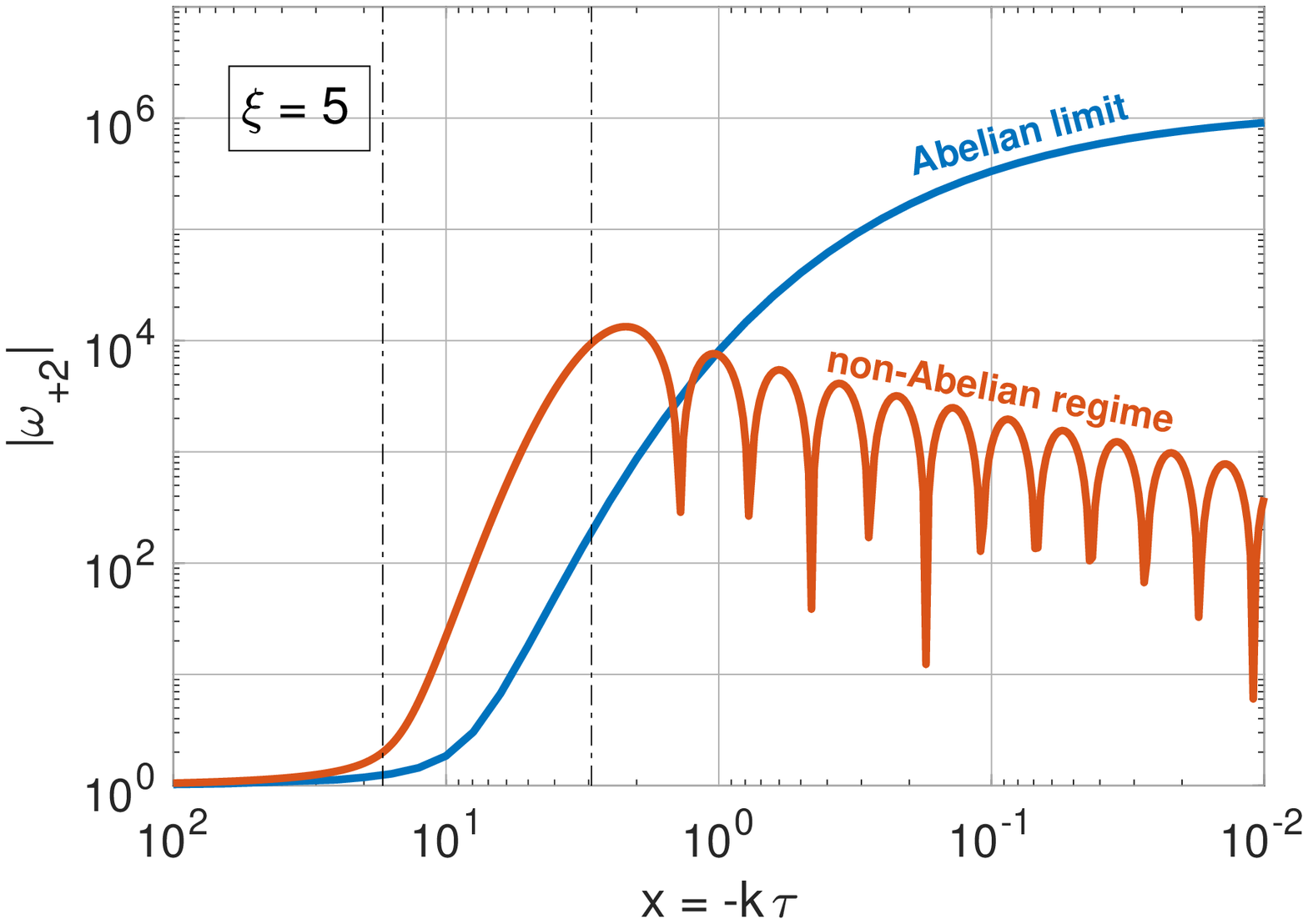}}
    \caption{ \textbf{Left}: Region for which the helicity $+2$ mode exhibits a tachyonic instability in the case of a background described by the $c_{2}$ solution. We also highlight our parameter example of $\xi = 5$ with the red horizontal line. \textbf{Right}:  Enhanced gauge field growth in the Abelian limit as well as non-Abelian regime. Additionally, we indicate with the two black vertical lines the estimated duration of the growth period in the non-Abelian regime, given by $\Delta x \simeq \xi\sqrt{8}$.
     }
    \label{fig:tachyonicexample}
\end{figure}

Remarkably, we also find a very good analytic approximation for the inherently non-Abelian case with $\xi \neq \text{const.}$ and $c_{i} = c_{2}$.
It has a similar form as before since the only change is encoded in the tachyonic mass term (now $\psi \neq 0$):
\begin{align}
\label{eq:omega+2nonabelian}
          \omega_{+2}^{k}(\tau) = \text{e}^{\kappa\pi/2}W_{-i\kappa,-i\mu}(2ik\tau)\bigg\rvert_{\xi = \xi \left( N_{h}(k) +{\Delta N} \right)}, 
\end{align}
where we defined $ \kappa \equiv (1+c_{2})\xi$ and $\mu \equiv \xi \sqrt{2 c_{2}-1/(4\xi^2)}$.
Here $N_h(k)$ labels the e-fold at which the mode $k$ crosses the horizon. The appearance of $\Delta N > 0$ indicates that the mode function is most sensitive to value of $\xi$ just before horizon crossing, when the tachyonic mass term is most relevant. In the entire parameter space of our interest, and for the scalar potential given in Eq.~\eqref{eq:scalar_potential}, we find $\Delta N = 3$ to be an excellent fit to the full result, obtained by numerically solving the coupled system of equations. Note that at early stages of inflation $\xi$ is approximately constant, and hence Eq.~\eqref{eq:omega+2nonabelian} smoothly matches to the Abelian result~\eqref{eq:variance_ab} after replacing $c_2 \mapsto 0$. 
Although both Abelian and non-Abelian solutions appear at the first sight to have very similar form, they behave completely different, see figure~\ref{fig:tachyonicexample}. 
This is because the background field presence suppresses the growth of the fluctuations -- which is of particular importance in our setup as we discuss in section~\ref{subsec:numeric}.
The main difference is that the region of tachyonic instability is bounded in the $(\xi,x)$ plane, leading to a saturation of the growth only shortly after it started.
This opens for us a viable window where we can calculate the linearized non-Abelian dynamics.

Let us take a step back and emphasize our viewpoint.
In the infinite past the unique physical vacuum state in de Sitter spacetime is of Bunch-Davies form.
This sets the initial conditions and directly translates into an absence of the gauge field background, such that it is described with the $c_{0}$ solution.
The gauge field fluctuations in this limit are well described by the pure Abelian case\footnote{Actually, in the limit of small gauge group coupling and/or small gauge field amplitude, any $SU(2)$ group will act like $m=2^2-1=3$ copies of an Abelian group. However, since the fields point in arbitrary directions, we take $m\equiv 1$ for simplicity.}.
This justifies, that we can in practice replace Eq.~(\ref{eq:su2fulleom}) and Eq.~(\ref{Hubble}) respectively with their Abelian limit in the Coulomb gauge according to\footnote{ The naive Abelian limit of the non-Abelian dynamics obtained by $g\to 0,~\psi\to 0$ just reflects that in the pure Abelian theory {the inflaton EOM does not receive any corrections at linear order in the gauge field fluctuations. We thus need to invoke the leading, second order contribution.}}
\begin{align}
\label{eq:eomabelianlimitphi}
- \frac{3g\alpha}{\pi f_\axion} \psi^2\left(\frac{\psi}{H} - \frac{\psi'}{H} \right) ~~ &\mapsto ~~ - \frac{\alpha}{\pi f_\axion} \frac{F_{\mu\nu}\widetilde{F}^{\mu\nu}}{4H^2}  \simeq - 2.4\cdot10^{-4} \frac{\alpha}{\pi f_\axion}  \frac{H^2}{\xi^4}e^{2\pi\xi}, \\
\label{eq:eomabelianlimithubble}
\frac{3}{2}H(\psi - \psi') + \frac{3}{2}g^2\psi^4 ~~ &\mapsto ~~ \frac{1}{4} \left( 4 F_{0\alpha}F^{\alpha 0} + F_{\mu\nu}F^{\mu\nu}  \right)\simeq 1.4\cdot10^{-4} \frac{H^4}{\xi^3}e^{2\pi\xi}.
\end{align}
The growth of $\xi \propto |a'|$ eventually triggers the transition to the inherently non-Abelian regime, manifesting itself through the $c_{2}$ background.
The time when this happens is given by Eq.~(\ref{eq:transition}) and we denote this as the matching point, as we reach the inherently non-Abelian phase.
This dynamic will basically last until inflation ends, i.e.\ until $\epsilon = (\ln H)' =1 $ is reached.
The time of the violation of the slow-roll condition in axion inflation with non-Abelian gauge fields defines $N=0$.

\begin{figure}[!h]
    \centering
    \subfigure{\includegraphics[width = 7cm]{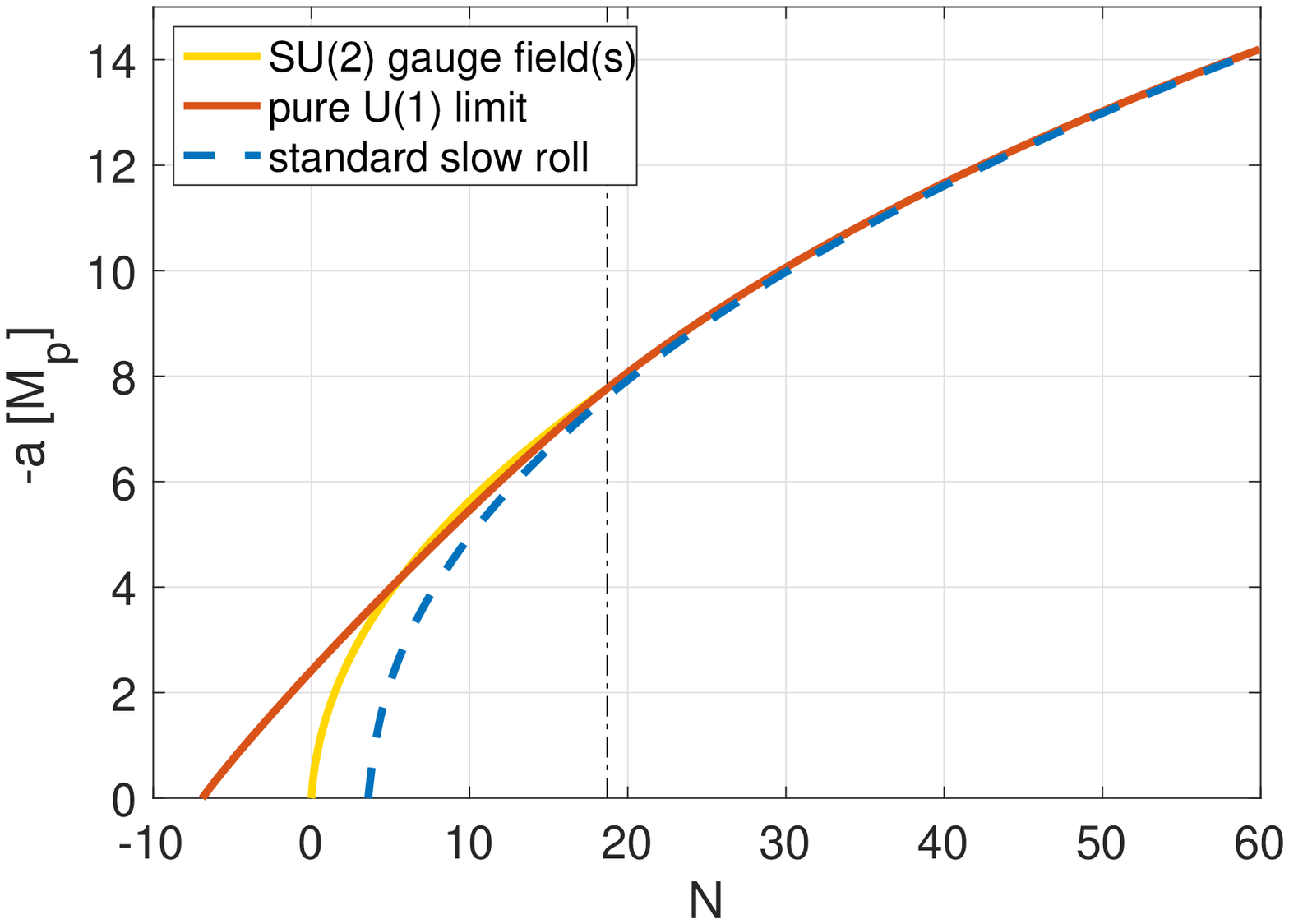}}
    \subfigure{\includegraphics[width = 7cm]{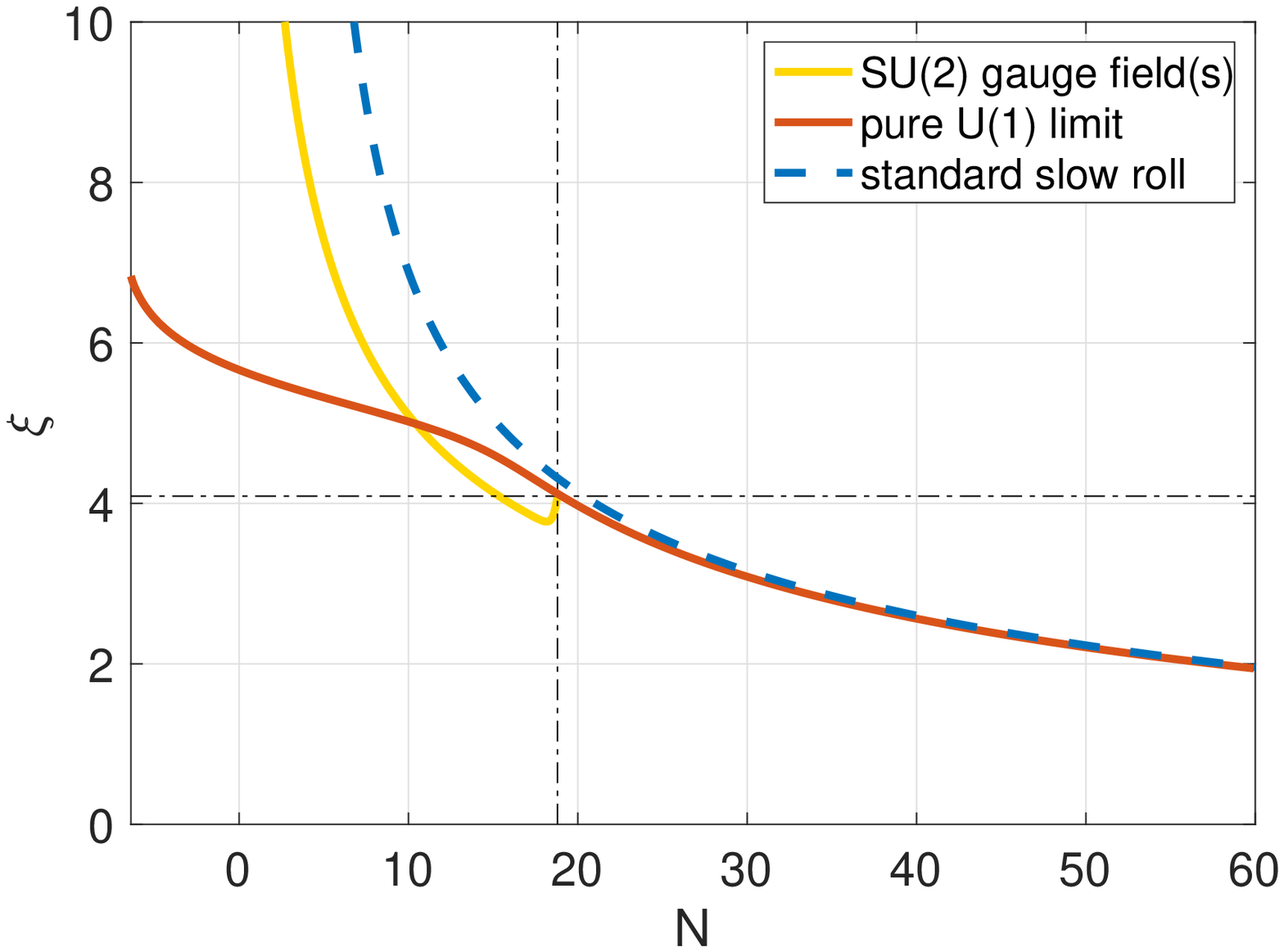}}
    \caption{ \textbf{Left}: Evolution of the axion for all three possible cases, i.e.\ (a) vacuum dynamics and in the presence of (b) only Abelian and (c) non-Abelian gauge groups. We choose the exemplary parameters $\Lambda^4 = 4.7 4\times 10^{-9}~M_{P}^4$ and $f_{\text{eff}} = 9.2~M_{P}$ for the potential of Eq.~(\ref{eq:scalar_potential}). Furthermore, the two couplings are fixed to {$\alpha/(\pi f_\axion) = 35$} and $g = 5\times 10^{-3}$.  The black vertical line indicates the matching point, obtained by solving Eq.~(\ref{eq:transition}). \textbf{Right}: Evolution of the $\xi$ parameter which encodes the complete information about the tachyonic instability in the gauge field sector of (b) and (c).
The matching point discontinuity of $\xi' \propto -\axion''$ is caused by the sudden change of the description explained in the main text.
    }
    \label{fig:exemplarypointstudy}
\end{figure}

Let us illuminate the setup with an exemplary parameter point study. 
We choose the axion scalar potential as given in Eq.~(\ref{eq:scalar_potential}) and the constants are given in the figure caption.
The resulting dynamics is shown in figure~\ref{fig:exemplarypointstudy}.

The rest of the paper will be dedicated to an in depth study of observational, numerical and theoretical constraints which this setup has to face.


\section{Constraints}
\label{sec:constraints}

Emerging chromo-natural inflation, as reviewed in the previous section, is subject to a number of restrictions arising from phenomenological constraints as well as from the consistency of the effective field theory (EFT) giving rise to Eq.~\eqref{eq:axion_inflation_lagrangian}. In addition, the linearized description in section~\ref{sec:axioninflation} requires perturbative control, which may not be guaranteed in the entire parameter space. This section is dedicated to discussing these limitations. {Their implications on the different regions of parameter space will be summarized in Sec.~\ref{sec:results}.}

\subsection{The effective field theory}
\label{sec:eftconstraints}

The most natural scalar potential for an axion-like particle is arguably given by
\begin{equation}
 V(a) = \Lambda^4 \left(1 - \cos\left(\frac{\axion}{f_\axion} \right) \right) \,,
 \label{eq:potential_naiv}
\end{equation}
with $f_\axion$ the fundamental axion decay constant appearing in Eq.~\eqref{eq:axion_inflation_lagrangian} and $\Lambda$ corresponding to the confinement scale of some additional dark sector non-Abelian gauge group. This is completely analogous to the non-perturbative mass generation for the QCD axion through instanton effects, and breaks the continuous $U(1)$ symmetry of the axion down to a discrete symmetry with periodicity $2 \pi f_\axion$.
However, as {it} is well known in the context of natural inflation~\cite{Freese:1990rb,Adams:1992bn} which is characterized by the scalar potential~\eqref{eq:potential_naiv}, slow-roll inflation in agreement with the Planck data~\cite{Akrami:2018odb} requires a periodicity scale $f_\axion$  of ${\cal O}(10) \,  M_P$.
This is problematic both from the view of a generic EFT\footnote{In the context of Peccei-Quinn axion models, $f_\axion$ is essentially the scale where the $U(1)$ Peccei-Quinn symmetry is spontaneously broken (see below). {Moreover}, we expect any global symmetry to be broken by (quantum) gravity effects above the Planck scale~\cite{Kallosh:1995hi}.}~\cite{ArkaniHamed:2003mz}
as well as from the point of view of string theory (which disfavours super-Planckian field excursions~\cite{Banks:2003sx}).
In the following, we discuss this problem and possible solutions first in a bottom-up approach based on a generic perturbative Peccei-Quinn model and then in a string theory inspired approach.

\subsubsection{Peccei-Quinn models \label{sec:PQ}}

\paragraph{Symmetry restoration.} During inflation, the Gibbons-Hawking temperature, $T_{dS} = H/(2 \pi)$, describes the thermal radiation inherent to de Sitter spacetime~\cite{Gibbons:1977mu}.
To avoid significant corrections due to unknown UV-physics entering above the cut-off scale $f_\axion$ introduced in Eq.~\eqref{eq:axion_inflation_lagrangian}, we must thus require\footnote{Alternatively, one may require that the typical quantum fluctuations of a scalar field in de Sitter spacetime, given by $H/(2 \pi)$, should be smaller than the cut-off scale. Numerically, this leads to the same condition.}
\begin{equation}
 \frac{H}{2 \pi} \ll  f_\axion \,.  
 \label{eq:eft1}
\end{equation}
This general argument can be made more explicit in the context of concrete axion models. The Peccei-Quinn (PQ) mechanism~\cite{Peccei:1977ur,Peccei:1977hh}, originally proposed to address the strong $CP$ problem of QCD, consists in adding a global $U(1)_{PQ}$ symmetry under which some (heavy) fermions are charged.
Below the PQ scale $T_{PQ}$ this symmetry is spontaneously broken by a complex scalar field $\Phi$ (charged under $U(1)_{PQ}$) obtaining a vacuum expectation value $v_{PQ}/\sqrt{2}$.
The angular degree of freedom of the resulting `mexican hat' potential is the shift-symmetric axion $\axion$. Non-perturbative effects may break the exact shift symmetry, leading to a $2 \pi v_{PQ}$ - periodic potential for $\axion$. The axion couples to the SM gauge fields as
\begin{equation}
 {\cal L} \supset c_{PQ} \frac{\axion}{ v_{PQ}} \frac{\alpha}{8 \pi} F_{\mu \nu} \widetilde{F}^{\mu \nu} \,,
 \label{eq:PQ}
\end{equation}
where $c_{PQ}$ is generically an order one number depending on the charge assignments of {heavy PQ fermions which have been integrated out in the effective description~\eqref{eq:axion_inflation_lagrangian}~\cite{Kim:1979if,Shifman:1979if}}.
Comparing to Eq.~\eqref{eq:axion_inflation_lagrangian}, we identify $f_\axion = 2 v_{PQ}/c_{PQ} \sim v_{PQ}$.

Above $T_{PQ}$ thermal effects restore the $U(1)_{PQ}$ symmetry and the complex PQ scalar $\Phi$ receives a large thermal mass. The low energy description~\eqref{eq:PQ} thus becomes completely inadequate, since the degrees of freedom in the symmetric phase are non-linearly connected to those of the broken phase.
The exact relation between $v_{PQ} \sim T_{PQ}$ depends on the scalar potential for $\Phi$.
In the original PQ model this is given by (see e.g.\ \cite{Kallosh:1995hi}),
\begin{equation}
 V_{PQ}(\Phi) = \lambda \left(|\Phi|^2 - \frac{v_{PQ}^2}{2} \right)^2 \,,
 \label{eq:VPQ}
\end{equation}
where $\lambda$ denotes a dimensionless coupling parameter.
At $\Phi = 0$, this leads to a tachyonic zero temperature mass term which receives thermal corrections in leading order from one loop self-interaction at high temperatures $T > |\Phi| $ \cite{Dolan:1973qd,Laine:2016hma},
\begin{equation}
  m^2_\Phi = - \lambda v_{PQ}^2 + 2\times\frac{\lambda}{6} T^2 \,,
\end{equation}
where the complex PQ field is normalized to have a canonical kinetic term.
The transition between the symmetric and the broken phase occurs at $m^2_\Phi = 0$.
Hence the effective description~\eqref{eq:PQ} (and correspondingly Eq.~\eqref{eq:axion_inflation_lagrangian}) is valid for
\begin{equation}
T_{dS} \ll T_{PQ} \quad \rightarrow \quad 
{\frac{H}{2 \pi} \ll f_\axion \,,}
 \label{eq:eft2}
\end{equation}
as anticipated in Eq.~\eqref{eq:eft1}.

\paragraph{The radial degree of freedom.} As discussed above, any axion model consists of (at least) two real degrees of freedom, contained in the complex scalar $\Phi$: the angular degree of freedom $\axion$ and the radial degree of freedom $\rho$. When discussing the EFT describing the axion $\axion$ (and in particular when assuming that it plays the role of the inflaton), we are implicitly assuming that the radial degree of freedom is significantly heavier and can be integrated out.
This is, of course, what we expect since the axion direction is {technically protected in the sense of t'Hooft~\cite{tHooft:1979rat}} by an approximate shift symmetry while the radial direction does not exhibit a symmetry protection.
Requiring the mass $m_a$ of the axion to be much lighter than the mass $m_\rho$ of the radial degree of freedom (given by Eq.~\eqref{eq:VPQ}) yields
\begin{equation}
 m_\axion^2 
 \ll m_\rho^2
 = 2 \lambda v_\text{PQ}^2 \qquad \rightarrow \quad f_\axion \gg  \frac{m_\axion}{\sqrt{2 \lambda}} \,.
 \label{eq:massbound}
\end{equation}
Assuming a perturbative realization of the PQ mechanism, $\lambda \leq 1$, Eq.~\eqref{eq:massbound} sets $f_\axion \gg m_\axion/\sqrt{2}$ as the lower bound for the axion decay constant.
The requirement to match the Planck data~\cite{Akrami:2018odb} with the (effective) axion scalar potential (see below) requires $m_\axion \sim 10^{-5}~M_P$.

\paragraph{Large field excursions.} The above points were general considerations about implementing cosmic inflation in PQ axion models. To match the observed value of the scalar power spectrum of the CMB anisotropies, we still need to address the problem of implementing super-Planckian field excursions.
{One possibility to achieve this is by considering not only a single axion, but $N$ axions $a^{(n)}$ with (for simplicity) similar fundamental axion decay constants, $f_{a}^{(n)} \sim f_\axion$. }
In `N-flation'~\cite{Dimopoulos:2005ac} (see also `assisted inflation'~\cite{Liddle:1998jc}), sub-Planckian field excursions of many individual axions lead to a large field excursion of the effective inflaton $\Delta a \sim \sqrt{N} \Delta a^{(n)}$.
To leading order, the resulting scalar potential for the lightest degree of freedom is just a quadratic potential and (in the regime $\Delta a \ll \sqrt{N} f_a$) can be modeled by Eq.~\eqref{eq:potential_naiv} after replacing $f_\axion \mapsto f_\text{eff} \sim \sqrt{N} f_\axion$.
The downside of this mechanism is that it requires a very large number of axions (though note that this can be reduced by allowing for kinetic mixing among the axions~\cite{Bachlechner:2014hsa}).
 In this sense, a more efficient multi-axion implementation is the KNP alignment mechanism~\cite{Kim:2004rp} which achieves $f_\text{eff} \gg f_\axion$ by appropriately adjusting the anomaly coefficients of the individual axion couplings to the hidden gauge sector.
 Note, however, that the probability to have all anomaly coefficients randomly at $\mathcal{O}(1)$ is $\mathcal{O}(f_{a}/f_{\text{eff}})$.
 But then already for small $N$, i.e. $N \log(N) \gtrsim 2 \log(f_{\text{eff}}/f_{a})$, this allows for effective axion decay constants $f_\text{eff} \sim \sqrt{N!} f_\axion$~\cite{Choi:2014rja}.
 The effective scalar potential for the lightest particle is again given by Eq.~\eqref{eq:potential_naiv} after replacing $f_\axion \mapsto f_\text{eff}$,
\begin{equation}
 V(\axion) = \Lambda^4 \left(1 - \cos\left(\frac{\axion}{f_\text{eff}} \right) \right) \,.
 \label{eq:scalar_potential}
\end{equation}
In the following, we will simply work with the effective potential~\eqref{eq:scalar_potential} with $f_\text{eff}$  of ${\cal O}(10) \,  M_P$, without specifying its UV origin.
The {lower bounds}
on the fundamental decay constant $f_\axion$ derived at the beginning of this subsection still apply, and correspond to upper bounds on the number of axions required, parametrized by $f_\text{eff}/f_a$.

\subsubsection{Axion monodromy \label{sec:monodromy}}

String theory generically predicts a large number of axions~\cite{Arvanitaki:2009fg}. Large field excursions are more difficult to obtain, but can be achieved by invoking multiple axions fields, as discussed above. In addition, axion monodromy~\cite{McAllister:2008hb,Silverstein:2008sg} provides a large field inflation model based on a single axion.
 Axion couplings to e.g.\ a D5-brane or D4-brane in the earliest type II string theory models~\cite{Silverstein:2008sg, Flauger:2009ab} or to a $4-$form field strength in $4$d effective descriptions~\cite{Kaloper:2008fb, Kaloper:2008qs, Kaloper:2011jz} ensure that the potential energy increases each time the axion is shifted by $2 \pi f_a$.
 Consequently the would-be periodic direction is `unwrapped' and typically receives a monomial potential. This allows for large field inflation while all flat directions (moduli) can be stabilized{, i.e.\ can be effectively integrated out}.
  We note that the non-perturbative moduli stabilization mechanisms are not captured in the EFT approach of Sec.~\ref{sec:PQ} and hence the bounds on $f_a$ derived there do not apply to the case of axion monodromy.
However, even employing monodromy, the control of backreaction typically limits the maximal field excursion achievable. The precise bound depends on the concrete realization but is typically in the ball-park $\Delta \axion/f_\axion \lesssim 10^3$, see e.g.~\cite{Silverstein:2008sg,Flauger:2009ab,McAllister:2016vzi}.

For the purpose of this paper, the scalar potential $V(\axion)$ is mainly a proxy for the time-evolution of the instability parameter $\xi$, governed by the inflaton velocity $\axion'$.
The parameter $\xi$ in turn controls the gauge field production which is at the core
of axion gauge field inflation,
as described in Sec.~\ref{sec:axioninflation}.
It was demonstrated in Ref.~\cite{Domcke:2016bkh} for the case of Abelian axion inflation that all monomial scalar potentials (including Eq.~\eqref{eq:scalar_potential} in the limit $a \ll f_\text{eff}$) fall into the same universality class.
This leads to the same predictions for all the phenomenology which is associated to the gauge field production.
The reason is that for all monomial potentials $V(\axion) \propto \axion^p$ in the weak backreaction regime, $\xi \propto \sqrt{\epsilon} \sim \sqrt{p/4}/\sqrt{N_{e}}$ with {$\epsilon \equiv a'^2/2$} denoting the first slow roll parameter.
This argument can be extended to the non-Abelian case discussed here, taking into account that the mild $p$-dependence of the proportionality factor can no longer be simply absorbed in the coupling $\alpha$.
Hence for simplicity, we will consider the effective potential~\eqref{eq:scalar_potential} also for the case of axion monodromy.
A more detailed discussion about implementing axion inflation in a string theory setup can be found e.g.\ in Ref.~\cite{Barnaby:2011qe} (for Abelian gauge fields) and Ref.~\cite{McDonough:2018xzh} (for non-Abelian gauge fields).

\subsection{CMB observations, gravitational waves and primordial black holes \label{sec:phenoprobes}}

We now turn to the phenomenological constraints on the scalar and tensor power spectra, both on large scales (relevant for the CMB) as well as on small scales (which are constrained by direct gravitational wave searches and constraints on primordial black holes).

\paragraph{CMB observations.} The coupling to gauge fields modifies the primordial perturbation spectra in a two ways: (i)~The gauge fields induce an additional effective friction which implies that the vacuum fluctuations contributing to the anisotropies in the CMB are sourced as the inflaton passes a different (lower lying) part of the scalar potential compared to the limit {$\alpha/f_{a} \rightarrow 0$~\cite{Anber:2009ua}. }This leads to a modification of the predictions for the amplitude of the scalar perturbations $A_s$, the scalar spectral index $n_s$ and the tensor-to-scalar ratio $r$, which characterize the scalar and tensor CMB anisotropies. (ii)~The gauge fields act as an additional classical source for scalar and tensor perturbations. These turn out to be highly non-gaussian, and thus constraints on the non-guassianity of the scalar perturbations severely constrain this contribution at the CMB scales~\cite{Barnaby:2011qe}.

As we will see in Sec.~\ref{sec:results}, in the context of emerging chromo-natural inflation, the CMB scales fall into the `Abelian limit' for the entire parameter space.
 In this case all constraints arising from point (ii) can be summarized in a bound on the parameter $\xi$ evaluated at CMB scales, $\xi_\text{CMB} \leq 2.5$~\cite{Barnaby:2011qe}.
The constraints arising from point (i) are more model dependent, since they depend on the amount of friction accumulated throughout the last 60 e-folds of inflation as well as on the details for the scalar potential.
A detailed discussion of this effect, including different classes of scalar potentials can be found in Ref.~\cite{Domcke:2016bkh}. Here, we will neglect this effect since by minor modifications, the scalar potential relevant for the CMB fluctuations can always be
adjusted
to reproduce acceptable values for $A_s$, $n_s$ and $r$.

\paragraph{Scalar and tensor power spectrum at small scales.} Since the inflaton velocity and hence the parameter $\xi$ typically increase during inflation, the gauge field production is generically more efficient towards the end of inflation. This leads to a strong enhancement of the scalar and tensor perturbation spectrum at scales much smaller than those probed by the CMB.

The tensor power spectrum is obtained by solving the coupled system of linear differential equations describing the helicity $+2$ component of the gauge field and metric perturbations (see e.g.\ Ref.~\cite{Domcke:2018rvv}). Denoting the Fourier coefficient of the positive helicity gravitational wave by $h_{+2}$ (normalized to $|h_{+2}| = 1$ in the far past), the gravitational wave spectrum is obtained as
\begin{equation}
 \Omega_\text{GW}(k) = \frac{\Omega_r}{24} \left( \frac{H}{2 \pi} \right)^2 \left( \frac{2}{M_P} \right)^2 \left[1 +  \left(k \tau |h_{+2}(k, \tau)| \right)^2 \right] \bigg|_{\tau = - 1/k}
\end{equation}
with {$\Omega_r = 9.12 \times 10^{-5}$}
denoting the present-day radiation energy density. Here the first term in the square brackets denotes the usual vacuum contribution (corresponding to the unenhanced $h_{-2}$
mode) whereas the second term accounts for metric perturbations sourced by the gauge fields. To express the gravitational wave spectrum in terms of the frequency of the present day GWs, note that a perturbation mode exiting the horizon at $N_{h}(k)$ e-folds before the end of inflation corresponds to a (comoving) frequency $f {= k/(2 \pi)}$ which is exponentially larger than the comoving mode $k_\text{CMB}$ exiting the horizon at $N_\text{CMB}$,
\begin{equation}
 \exp(N_\text{CMB} - N_{h}) = \frac{2 \pi f}{k_\text{CMB}} \,.
\end{equation}

Similarly, the scalar power spectrum receives contributions from vacuum fluctuations as well as a contribution sourced by the enhanced helicity $+2$ gauge field fluctuations. Here, due to helicity conservation, the dominant contribution arises to second order in $w_{+2}$ and can be estimated as~\cite{Domcke:2018rvv} (see also \cite{Papageorgiou:2018rfx})
\begin{equation}
 \left( \Delta_s^2 \right)^\text{2nd} \sim \left( \frac{\delta \axion}{a'} \right)^2 \qquad \text{with} \quad \delta \axion = -{ \frac{\alpha}{12 \pi f_\axion H^2}} \delta( \langle F \widetilde{F} \rangle) \,.
\end{equation}

Fluctuations at small scales are observationally much less constrained than the length scales relevant for the CMB anisotropies. The scalar perturbation spectrum is bounded by the requirement of not overproducing primordial black holes in the gravitational collapse of overdense region after horizon re-entry~\cite{Linde:2012bt}, $A_s \lesssim 10^{-4..-2}$, with the exact value depending on the details of the non-gaussian statistics.  The most stringent constraint on the tensor power spectrum arises from the non-observation of a stochastic gravitational wave background in LIGO~\cite{Abbott:2017xzg}, $\Omega_\text{GW}(30~\text{Hz}) \lesssim 10^{-7}$.
However, we verified numerically that neither of these two conditions impose any additional constraints on the parameter space which remains open after imposing the constraints discussed in Sec.~\ref{sec:results}.
In particular, despite the strong enhancement of the scalar perturbations arising from non-linear interactions (compared to the linearized result), c.f.\ Refs.~\cite{Domcke:2018rvv,Papageorgiou:2018rfx}, the primordial black hole production remains irrelevant in the parameter space in question.

\subsection{Non-linear interactions}
\label{subsec:numeric}

{In the last part of this section we turn to technical limitations of our analysis. Throughout this work we employ linear perturbation theory for the gauge field fluctuations. In the following we summarize the limitations of this framework.}

\paragraph{Abelian limit.} In the Abelian limit, the linearization is justified if the non-Abelian interactions are negligible, i.e.\ $ g A^2 \ll \partial_\mu A$. Inserting Eq.~\eqref{eq:variance_ab} this implies for the Fourier mode $k$,
\begin{equation}
0.008 \times \exp(2.8 \xi) \ll x/g \,.
\label{eq:pert_abelian}
\end{equation}
At any given time, the dominant contribution to the gauge field spectrum in the Abelian limit arises from roughly horizon-sized modes, and we thus require Eq.~\eqref{eq:pert_abelian} to hold for $x = 1$.
Given the rapid growth of $\langle A^2 \rangle$ with $\xi$, this roughly coincides with the `matching point' defined in Eq.~\eqref{eq:transition}.
Thus it is ensured by construction that non-linear interactions can be neglected in the Abelian limit.
Very close to the matching point this approximation becomes less accurate,
contributing to the systematic uncertainties of the matching procedure.

\paragraph{Matching procedure.} When switching from the Abelian to the non-Abelian description, our modeling induces a discrete change in the friction term of the equation of motion for the classical inflaton field, see Eq.~\eqref{eq:eomabelianlimitphi}. As long as this friction term is small compared to the Hubble friction~\cite{Dimastrogiovanni:2012ew},
\begin{equation}
 {\frac{\alpha \xi H}{g \pi f_\axion} \ll 1 \,,}
 \label{eq:magnetic_drift}
\end{equation}
this discrete change only induces a small perturbation and after a few e-folds the system reaches an equilibrium configuration. By excluding the regime of a few e-folds around the matching point from our phenomenological analysis, this systematic uncertainty is well taken into account. On the other hand, in the regime where the gauge friction dominates over the Hubble friction (referred to as `magnetic drift' of `strong backreaction' regime~\cite{Adshead:2012kp,Dimastrogiovanni:2012st,Dimastrogiovanni:2012ew,Adshead:2013qp}), the matching procedure described in Ref.~\cite{Domcke:2018rvv} comes with significant systematic uncertainties (for more details, see appendix~\ref{app:approximation}).
Imposing Eq.~\eqref{eq:eft1}, we see that $g < 2\pi /\xi$ is a sufficient condition to satisfy Eq.~\eqref{eq:magnetic_drift}. Inserting the matching condition~\eqref{eq:transition}, we note that in the entire regime of validity of the effective field theory, we are ensured to be in the weak backreaction regime. {Similarly, considering viable axion monodromy with $\Delta \axion/f_\axion \lesssim 10^3$ (see Sec.~\ref{sec:monodromy}) enforces the weak backreaction regime.}
Consequently, in this regime the systematic uncertainties associated with the matching procedure in the inflaton equation of motion are under control.

\paragraph{Non-Abelian regime.} In the non-Abelian regime, the background gauge field induces an effective mass for the gauge field fluctuations. However, as shown in Eq.~\eqref{eq:ode+2mode}, one of the gauge field degrees of freedom retains a tachyonic mass term for a certain range of Fourier modes at any given time. The resulting growth of these fluctuations, see Eq.~\eqref{eq:omega+2nonabelian}, in combination with the monotonic growth of the instability parameter $\xi$, will eventually lead to the breakdown of the linearized description {for the gauge field fluctuations} at $\langle \omega_{+2}^2 \rangle^{1/2} \sim R \psi$
At this point, there is no fundamental issue: we simply enter the inherently non-linear regime of the non-Abelian field theory. However, the tools employed here are inadequate to describe this regime. A lattice simulation of this system could resolve these non-linearities, however the implementation in an exponentially expanding universe is notoriously hard and state-of-the-art simulation only manage to cover a few e-folds, even for Abelian field theories~\cite{Adshead:2015pva,Cuissa:2018oiw}. Consequently, we are forced to limit our analysis to the regime of small fluctuations around the gauge field background.

When imposing this constraint, we note that none of {the} phenomenological constraints listed in Sec.~\ref{sec:phenoprobes} has any significant sensitivity below $N_{e} \sim 10$. The main impact of non-linear interactions in these last few e-folds of inflation is thus to potentially shift the end point of inflation with respect to the estimate obtained in the linearized theory. This would shift the reference scale at which any observables (i.e. CMB observables or direct GW searches) probe the scalar and tensor power spectra. Since however the total amount of e-folds from the CMB scales to the end of inflation is in any case subject to uncertainties related to the details of the reheating epoch {(see e.g. Ref.~\cite{Liddle:2003as})}, we will in the following ignore any violation of the linearization condition if it occurs at $N_{e}\lesssim 10$.
Estimating the amplitude of the fluctuations in the non-Abelian regime with Eq.~(\ref{eq:omega+2nonabelian}),
\begin{align}
\label{eq:nonabelianfluctuation}
   \langle \delta A_{\text{NA}}^{2} \rangle (N_{h})
   = \int \frac{\text{d}^3k}{(2\pi)^3} \, \frac{|\omega_{+2}^{k}(\tau(N_{h}))|^2}{2k}
    \simeq \frac{1}{-\tau(N_{h})^2}\int \frac{\text{d}x \, x}{2\pi^2} |\omega^k_{+2}(x)|^2 \bigg|_{\xi = \xi(N_{h}(k)+3)}\,,
\end{align}
leads to the following criterion {for which the linearized analysis is justified},
\begin{equation}
    \sqrt{\langle \delta A_{\text{NA}}^{2} \rangle} \ll R\psi \quad \text{or} \quad
   -\tau g \sqrt{\langle A_{\text{AB}}^2} \rangle \leq c_{2} \xi \, \big\rvert_{N_{e} \geq 10} \,.
   \label{eq:strong}
\end{equation}
The second half of this criterion corresponds to the situation where the matching point between the Abelian and non-Abelian regime is found to be only at $N_{e} \leq 10$.
This excludes by construction the possibility that non-Abelian effects become dominant at $N_{e} \geq 10$.

\paragraph{Scalar perturbations.} It was pointed out in Ref.~\cite{Dimastrogiovanni:2012ew} that the scalar perturbations feature an instability for $2 < \xi < 2.12$ which destroys the homogeneity of the inflaton field (see also Ref.~\cite{Domcke:2018rvv}).
This can lead to a premature end of inflation.
 Inserting Eq.~\eqref{eq:variance_ab} into Eq.~\eqref{eq:transition}, we note that the non-Abelian regime is typically characterized by much larger values of $\xi$. A matching point of $\xi < 2.12$ is only obtained for $g \gtrsim 1$. We hence conclude that the possibility of an instability in the scalar sector can be excluded for perturbative gauge couplings in the setup of emerging chromo-natural inflation.

\paragraph{Comparison with usual CNI setup.}
Before concluding this section, let us briefly comment on the more standard chromo-natural inflation setup, in which the non-trivial homogeneous and isotropic gauge field background is taken as an initial condition, present already when the CMB scales exit the horizon.
The constraints related to the scalar potential of the axion, discussed in Sec.~\ref{sec:eftconstraints}, are essentially insensitive to this distinction since they are intrinsic to the axion sector.
On the contrary, the predictions for the scalar and tensor power spectra (Sec.~\ref{sec:phenoprobes}), in particular at large scales are very different. The possibility that the non-trivial homogeneous and isotropic vacuum configuration exists already at the time the CMB scales exited the horizon is severely constrained by current CMB observations, to the point of excluding a scalar potential as in Eq.~\eqref{eq:scalar_potential}~\cite{Dimastrogiovanni:2012st,Dimastrogiovanni:2012ew,Adshead:2013qp,Adshead:2013nka}. Finally, the impact of non-linear gauge field fluctuations discussed in Sec.~\ref{subsec:numeric} is also sensitive to this choice of initial condition for the background gauge field. The conclusion is however similar, i.e.\ non-linear interactions were found to be not problematic in the magnetic drift regime~\cite{Adshead:2016omu}.


\section{Results and Discussion}
\label{sec:results}

\paragraph{The parameter space of emerging chromo-natural inflation.} Fixing the parameters of the scalar potential $f_\text{eff} = 9.2~M_P$ and $\Lambda = 8.3 \times 10^{-3}~M_P$ as suggested by the Planck data\footnote{{In fact, Eq.~\eqref{eq:scalar_potential} is only marginally compatible with the latest Planck data at the $2 \sigma$ level~\cite{Akrami:2018odb}. Here, we shall not be concerned with this tension, since it can be remedied by minor modifications of the scalar potential which do not significantly affect our main results.}}, there are only two remaining free parameters in the model: the gauge coupling $g$ and the effective coupling between the axion and the gauge fields, $\alpha/f_\axion$ with $\alpha = g^2/(4 \pi)$\footnote{Note that actually the two fundamental free parameters are the gauge coupling $g$ and the fundamental axion decay constant $f_{a}$. When we plot in the $(g,\alpha/f_{a})$ plane the axes are thus related. However, we decide for reasons of clarity and convenience not to plot in the $(g,f_{a})$ plane. The conclusions remain, of course, the same. }. 
Our discussion will be limited to perturbative gauge couplings $g \leq 1$. If the CMB scales exit the horizon in the Abelian limit (which we will find to be the case in the entire parameter space), the non-observation of non-gaussianities in the CMB scalar power spectrum limits $\alpha/(\pi f_\axion) \lesssim 35$~\cite{Barnaby:2011qe}.
On the other hand, very small values of the coupling parameters $g$ and $\alpha/f_\axion$ simply imply a decoupling of the gauge field sector and we recover standard single field slow-roll inflation.  Our study is complementary to the recent work~\cite{Maleknejad:2018nxz}, which studies the parameter space of different types of non-Abelian axion inflation models under the assumption that the non-trivial isotropic gauge field background is present already when the CMB scales exited the horizon.

\begin{figure}[!h]
\center
 \includegraphics[width = 0.75 \textwidth]{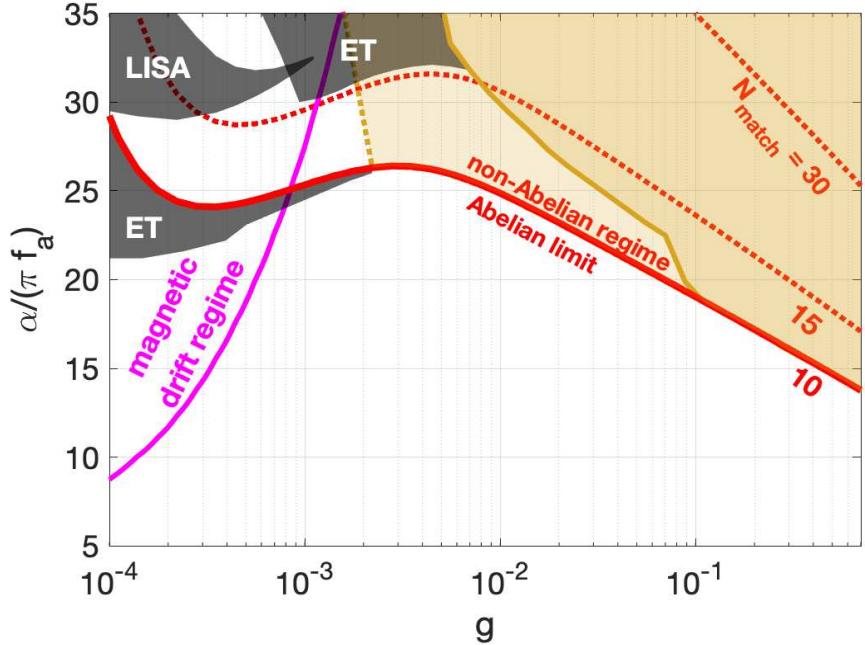}
 \caption{Parameter space of emerging chromo-natural inflation. The red lines indicate contour lines counting the number of e-folds in the non-Abelian regime, i.e.\ from the matching point to the end of inflation. The shaded yellow region marks when Eq.~\eqref{eq:strong} is violated, indicating the need of non-perturbative methods. The magnetic drift regime typically discussed in the CNI literature~\cite{Adshead:2012kp,Dimastrogiovanni:2012st,Dimastrogiovanni:2012ew,Adshead:2013qp} is to the left of the magenta line. The black shaded regions indicate a GW signal within the range of the Einstein Telescope and LISA.}
 \label{fig:pheno}
\end{figure}

In Fig.~\ref{fig:pheno} we display the qualitative different regions of emerging chromo-natural inflation in this parameter space. 
The matching condition~\eqref{eq:transition} between the Abelian and non-Abelian regime depends on the gauge coupling $g$ and the effective instability parameter $\xi$, which in turn is proportional to $\alpha/f_\axion$. 
This is reflected by the red contours in Fig.~\ref{fig:pheno}, counting the number of e-folds from the matching point to the end of inflation. 
In our analysis, a special role is played by $N_\text{match} = 10$ (indicated by the solid red line), since for $N_\text{match} < 10$ the gauge fields effectively behave like Abelian gauge fields in the phenomenologically relevant regime ($N_{e} \gtrsim 10$). 
For $N_\text{match} > 10$ the requirement of perturbativity  becomes non-trivial, and indeed we find Eq.~\eqref{eq:strong} to be violated in the shaded yellow region above the solid yellow line. This regime requires a non-linear treatment (e.g.\ a lattice simulation), prohibiting us from making any predictions with the linear formalism applied in this work.
 In the lighter yellow shaded region to the right of the dashed yellow line we find $ -\tau H \langle \delta A_{\text{NA}}^{2} \rangle^{1/2}/\psi > 1/10$, 
 implying that perturbativity is marginally fulfilled at best (note that in the linearization of Eq.~\eqref{eq:axion_inflation_lagrangian} we have dropped ${\cal O}(10)$ non-linear terms).\footnote{Note that Ref.~\cite{Maleknejad:2018nxz} introduces a more refined criterion based on the effect of the induced current on the gauge field background EOM. This corresponds to re-summing all non-linear terms and hence (as we have verified explicitly numerically), the condition $\langle \delta A_\text{NA}^2 \rangle^{1/2}/(R \psi) = 1/10$ corresponds to an ${\cal O}(1)$ backreaction from the induced current.} The results obtained in this regime should thus be taken with a big grain of salt.

The gray shaded regions indicate a GW signal within the reach of the future Einstein Telescope~\cite{ETdesignreport} and LISA~\cite{Caprini:2015zlo}. 
Since the stochastic gravitational wave background sourced by the gauge fields is maximally chiral~\cite{Cook:2011hg,Dimastrogiovanni:2012ew,Adshead:2013qp}, this could be a smoking gun signal for axion inflation~\cite{Seto:2006hf,Smith:2016jqs}. 
In particular, for the Einstein Telescope the lower gray region indicates a GW signal arising from the Abelian limit of the theory, i.e.\ when the the peak sensitivity of the Einstein telescope corresponds to frequencies below $f(N_\text{match})$, where $f(N_\text{match})$ denotes the frequency of the GWs exiting the horizon at $N_\text{match}$. Conversely, the upper gray region corresponds to GW signals from the inherently non-Abelian regime. 
For $15 \lesssim N_\text{match} \lesssim 10$, the peak sensitivity of the Einstein Telescope falls too close to the matching point, and the systematic uncertainties of our procedure prevent us from making a reliable prediction for the GW amplitude.
On the other hand, we can only predict a GW signal detectable by LISA from the Abelian limit of the theory.
As the peak sensitivity is in the range  $20 \lesssim N_{e} \lesssim 25$ , a signal from the inherent non-Abelian dynamics can not be reliably predicted because the linearized approach is not justified anymore (c.f. the $N_{\text{match}} = 30$ contour line in Fig.~\ref{fig:pheno}).
Both areas for possible GW detection arising from the Abelian limit probe similar values of the gauge coupling $g$, but a signal from either Einstein Telescope or LISA would point to a higher or lower fundamental axion decay constant respectively (in the range of $ 5 \times 10^{-12}~M_P \lesssim f_{a} \lesssim 4 \times 10^{-9}~M_P$).

Finally, the magenta line indicates the `magnetic drift' or `strong backreaction' regime  commonly studied in the literature, see e.g.\ \cite{Adshead:2012kp,Dimastrogiovanni:2012st,Dimastrogiovanni:2012ew,Adshead:2013qp}. {To obtain this contour, we have checked numerically for deviations of the axion background evolution from the usual Hubble friction dominated solution. 
It agrees roughly with the estimate given in Eq.~\eqref{eq:magnetic_drift}. 
Note that since $\alpha = g^2/(4 \pi)$ the magenta curve gives a lower bound on $g$ for fixed $f_\axion$ (and vice versa an upper bound on $f_{a}$ for fixed $g$).
Most studies of the magnetic drift regime assume the existence of the non-trivial homogeneous and isotropic $c_2$-solution from the onset of the observable part of inflation, i.e.\ for all $N_{e} \lesssim 60$. However, Fig.~\ref{fig:pheno} illustrates that this situation does not emerge dynamically when starting from the Bunch-Davies vacuum. The entire magnetic drift regime is characterized by $N_\text{match} < 30$, indicating that only during the last 30 e-folds of inflation (at most) the Abelian fluctuations become large enough to trigger a non-trivial homogeneous and isotropic gauge field configuration with a sizeable probability. In particular, to good approximation $\psi = 0$ when the scales relevant for the CMB exited the horizon. The CMB fluctuations thus closely resembles those generated in Abelian axion inflation, which in turn for $\alpha/(\pi f_\axion) \lesssim 35$ closely resemble those of single field slow-roll inflation, in agreement with current observations.

\paragraph{EFT constraints.} Fig.~\ref{fig:eft} visualizes the constraints on the effective field theory of axion gauge field inflation as discussed in \ref{sec:eftconstraints}.
\begin{figure}[!h]
\center
 \includegraphics[width = 0.75 \textwidth]{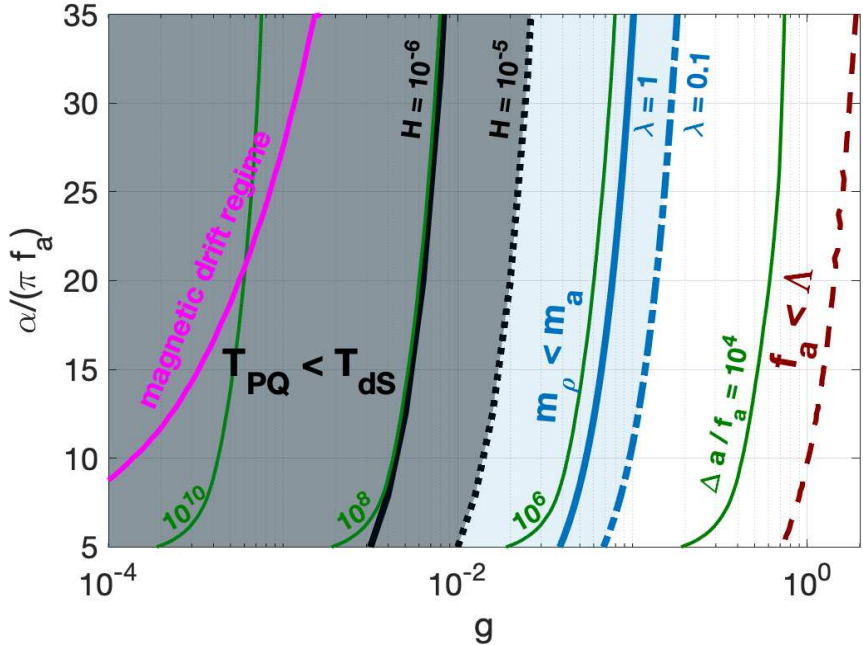}
 \caption{EFT constraints on chromo-natural inflation. The gray shaded region is due to the restoration of the Peccei-Quinn symmetry during inflation. In the blue region, the radial degree of freedom $\rho$ of the Peccei-Quinn field cannot be integrated out. In the remaining region, the thin green contours indicate the ratio $\Delta \axion/f_\axion$, corresponding to the degree of alignment/winding required to obtain a sufficiently large effective field excursion. To the left of the dashed red line the explicit breaking of the Peccei-Quinn symmetry occurs before the spontaneous breaking.}
 \label{fig:eft}
\end{figure}
 Interpreting the inflaton $\axion$ as the angular degree of freedom of a perturbative Peccei-Quinn symmetry (as in Sec.~\ref{sec:PQ}) excludes the shaded regions (see below). 
These bounds do a priori not apply if the moduli space is subject to non-perturbative stabilization,
as occurs e.g.\ in axion monodromy (see Sec.~\ref{sec:monodromy}).
However, in this case a similar region of the parameter space is disfavoured since it requires the field excursion $\Delta \axion$ during inflation to be many orders of magnitude larger than the fundamental axion decay constant $f_\axion$. 
This is indicated by the green contour lines in Fig.~\ref{fig:eft}, where we recall that values of $\Delta a/f_a \gg 10^3$ have proven to be very difficult to implement in concrete models.

Let us discuss in some more detail the restrictions for a Peccei-Quinn type axion. The requirement that the Peccei-Quinn symmetry associated with the axion $\axion$ is spontaneously broken during inflation (see Eq.~\eqref{eq:eft1}) excludes the gray region to the left of the black contours which indicate the corresponding Hubble scale during inflation. For the scalar potential~\eqref{eq:scalar_potential}, the Hubble rate varies from about $10^{-5}~M_P$ to $10^{-6}~M_P$ over the course of the last 60 e-folds of inflation.
A more stringent constraint can be derived by requiring that the axion $\axion$ is the lightest particle in the spectrum, and in particular lighter than the radial degree $\rho$ of the complex Peccei-Quinn scalar. This is indicated by the blue shaded region in Fig.~\ref{fig:eft} for different values of the quartic coupling of the Peccei-Quinn field. We see that perturbative realizations of the Peccei-Quinn mechanism $(\lambda \leq 1)$ are incompatible with gauge couplings smaller than a few times $10^{-2}$. For smaller values of $\lambda$, this bound only becomes stronger as indicated by the dashed contour labeled $\lambda = 0.1$.

Finally, for reference, the dashed red contour corresponds to $f_\axion = \Lambda$, i.e.\ only to the right of this contour the spontaneous breaking of the Peccei-Quinn symmetry occurs before the explicit breaking due to instanton effects.
We see that for essentially the whole perturbative parameter space, there is no time in cosmic history in which the shift-symmetry of the axion is unbroken.
This is the reverse order of $f_\axion$ and $\Lambda$ than one than is usually familiar with in the context of e.g.\ the QCD axion and natural inflation~\cite{Freese:1990rb}.
This may be relevant when discussing the naturalness of the initial conditions required for the axion field, however a detailed discussion of this question is beyond the scope of the present paper.

\paragraph{Discussion.} We summarize the most important constraints from the above discussion in Fig.~\ref{fig:combined}. 
\begin{figure}[!h]
\center
 \includegraphics[width = 0.75 \textwidth]{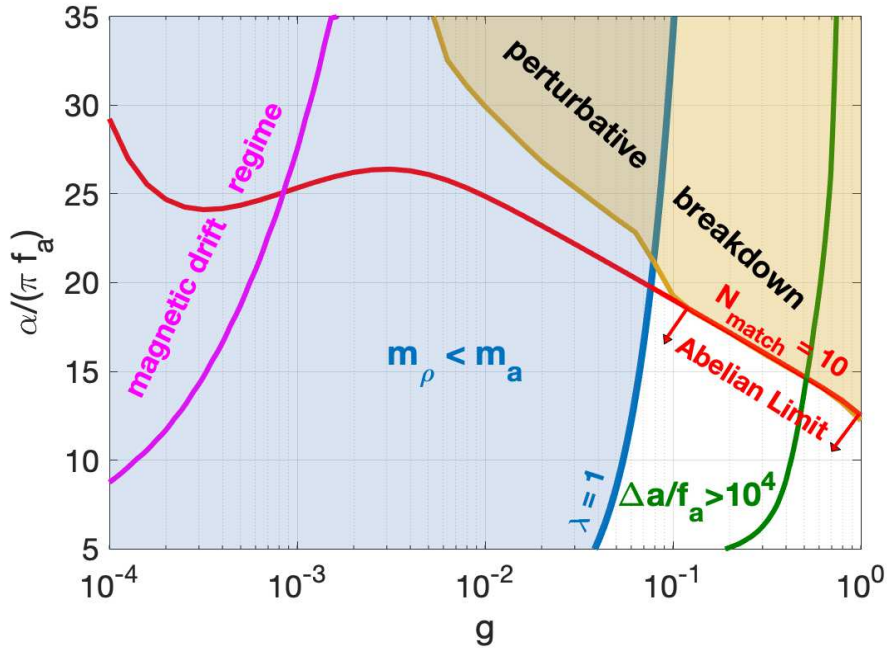}
 \caption{Summary of the viable parameter space for emerging chromo-natural inflation. Color-coding as in Figs.~\ref{fig:pheno} and~\ref{fig:eft}.
 }
 \label{fig:combined}
\end{figure}
In the context of a Peccei-Quinn type axion model, a large part of the parameter space (shaded blue region) is in conflict with basic requirements for a consistent EFT description of axion inflation. In the regime of $m_\rho \sim m_\axion$, it may nevertheless be possible to find a consistent inflation model involving a dynamical axion field, however this will necessarily be a two-field model involving both the angular and the radial degree of freedom. For $m_\rho \ll m_\axion$, the axion field will quickly become stabilized in a local minimum, and the gauge field production induced by $\axion' \neq 0$, which is at the heart of 
axion gauge field inflation,
will become irrelevant. 
{In string theoretical realizations, such as in axion monodromy, the invoked non-pertubative stabilization of the moduli space does not improve this situation.}
The reason is that the phenomenologically interesting parameter space pushes the fundamental axion decay constant $f_{\axion}$ to rather small values,
which together with the CMB requirement of a large field excursion for monomial potentials leads to extremely large winding numbers.

On the contrary, the yellow shaded region labeled `perturbative breakdown' does not indicate any fundamental inconsistency of the theory, but simply the inability to derive results in the linearized theory. 
A study of the full non-linear dynamics is however beyond the scope of this paper. 
We hope that this work may motivate a closer investigation of this regime using e.g. lattice simulations.
 Being inherently non-Abelian, we expect a direct coupling between gauge field perturbations and gravitational waves, as in the original CNI proposal~\cite{Dimastrogiovanni:2012ew,Adshead:2013qp}.
 Together with the large effective coupling $\alpha/f_\axion$ and the relatively large values of $N_\text{match}$, this regime is the most promising for the observation of gravitational waves associated with the non-Abelian nature of the gauge fields in direct gravitational wave detectors such as LIGO, LISA or the Einstein Telescope.

We confirm that the magnetic drift regime can be consistently described within the linearized theory. However, we find this regime to be highly problematic in the context of the UV completions considered. Moreover, starting from Bunch-Davies initial conditions for the gauge fields in the far past, we find that the gauge field fluctuations are not large enough to trigger the non-trivial background solution for the gauge field by the time the CMB scales exit the horizon. An implementation of the magnetic drift regime already at $N_{e} \sim 60$, as commonly studied in the literature, thus requires some non-trivial construction to avoid the constraints discussed in Sec.~\ref{sec:eftconstraints}, as well as some other mechanism to set the required initial conditions for the background gauge field.

We note that in terms of the fundamental model parameters $f_\axion$ and $g$, the constraints discussed in Sec.~\ref{sec:eftconstraints} are simply lower bounds on $f_\axion$, essentially independent of $g$. Being within the magnetic drift regime implies a lower bound on $g/f_\axion$. The non-observation of non-gaussianity in the CMB spectrum yields an upper bound on $g^2/f_\axion$ (recall that we find the horizon exit of the CMB scales to fall within the Abelian limit) which pushes the viable magnetic drift regime to $g \lesssim 10^{-3}$. The breakdown of the linearized analysis mainly effects large values of $g$.


\section{{Connecting to the Standard Model}}
\label{sec:su2u1}

In the previous sections, we discussed a setup of axion inflation coupled to $SU(2)$ non-Abelian gauge fields through a Chern-Simons term.
The derived constraints, {together with the requirement of reaching the intrinsically non-Abelian regime, push} the model to large gauge group couplings and point towards the necessity of an explicit evaluation of the non-linear interactions.
We now present a possibly more realistic scenario of axion interactions simultaneously with Abelian and non-Abelian gauge fields.
Such a scenario is indeed desirable if we ultimately couple the dark inflation sector to the Standard Model to successfully reheat our Universe.

We consider the Lagrangian
\begin{align}
\label{eq:unificationlagrangian}
\mathcal{L}= - \sqrt{-|g_{\mu\nu}|}\left( \frac{1}{2}(\partial_{\mu}\axion)^2 + V(\axion) + \frac{1}{4} F_{Y,\mu\nu}F_{Y}^{\mu\nu} + {\frac{\alpha_{U(1)}}{4 \pi f_\axion} }\axion F_{Y,\mu\nu}\widetilde{F}_{Y}^{\mu\nu}  + \frac{1}{4} F_{W,\mu\nu}F_{W}^{\mu\nu} + { \frac{\alpha_{SU(2)}}{4\pi f_{a}} }a F_{W,\mu\nu} \widetilde{F}_{W}^{\mu\nu}  \right),
\end{align}
where we change the notation to make the connection to the SM obvious and denote the non-Abelian field strength as $F_{W,\mu\nu}$ and the Abelian one as $F_{Y,\mu\nu}$.
The resulting EOM for the Abelian gauge fields reads
\begin{equation}
 \partial_{\mu} F_{Y}^{\mu\nu} + {\frac{\alpha_{U(1)}}{\pi f_{a}}}\partial_{\mu}\left( a \sqrt{-g}\widetilde{F}_{Y}^{\mu\nu} \right) + \partial^{\nu} \partial_{\mu} A_{Y}^{\mu} = 0 \,, 
\end{equation}
while the EOM for the homogeneous non-Abelian gauge field and inflaton component read
\begin{align}
     \psi'' - 3  \psi' \left(1 - \frac{H'}{3H} \right) + \psi \left(2 - \frac{H'}{H} \right) + 2g^2\frac{\psi^3}{H^2} &=  {\frac{g\alpha_{SU(2)}}{\pi f_\axion} }\psi^2 \frac{\axion'}{H},\\
     \axion'' - 3\axion' \left( 1 - \frac{H'}{3H} \right) + \frac{\partial_{\axion}V(\axion)}{H^2} &=  {\frac{3g\alpha_{SU(2)}}{\pi f_\axion}}\psi^2\left(\frac{\psi}{H} - \frac{\psi'}{H} \right) - { \frac{\alpha_{U(1)}}{4\pi f_{a}} }\sqrt{-g} F_{Y,\mu\nu}\widetilde{F}_{Y}^{\mu\nu}.
     \label{eq:combined}
\end{align}
For the Abelian gauge field the choice of Coulomb gauge and the decomposition into its Fourier modes leads to the analytical result (as given in Eq.~(\ref{eq:abelmode})),
\begin{align}
A_{Y,+}(k,\tau) = \frac{1}{\sqrt{2k}}\text{e}^{\pi\xi/2} W_{-i\xi,1/2}(2ik\tau)\,.
\end{align}
There are now essentially two options to circumvent the afore derived constraints.
The first one is to effectively decouple the non-Abelian gauge fields from the dynamics by demanding $\alpha_{SU(2)} \ll \alpha_{U(1)}$, while still accounting for the non-gaussianity constraint $\alpha_{U(1)}/(\pi f_{a}) \lesssim 35$.
This allows to pair a large value of the axion decay constant $f_\axion$ (as indicated by the constraints in Sec.~\ref{sec:eftconstraints}) with a large value of the $U(1)$ gauge coupling, so as to reach the phenomenologically interesting regime of large $\alpha_{U(1)}/ f_\axion$ without having to deal with significant non-linear gauge field interactions, which are controlled by the $SU(2)$ gauge coupling. In this case the phenomenology will be essentially indistinguishable from Abelian axion gauge field inflation.

The second option is to note that for $\alpha_{SU(2)} \simeq \alpha_{U(1)}$, as holds in the SM at high energies, the backreaction on the inflaton dynamics in Eq.~\eqref{eq:combined} is dominated by the Abelian gauge fields. In the regime of validity of the linearized analysis, this is easily confirmed: the background field energy density in the non-Abelian sector grows proportional to the fourth power in $\xi$, whereas for the gauge field in the Abelian sector we have an exponential dependence on $\xi$. The latter hence quickly comes to dominate the backreaction. This is also visible in figure~\ref{fig:exemplarypointstudy}, where we see that the coupling to Abelian gauge fields leads to a strong deviation of the inflaton velocity from the standard slow-roll Hubble friction scenario, whereas the same coupling to non-Abelian gauge fields closely follows the standard slow-roll solution. 
We expect a similar behaviour to hold even beyond the linearized analysis, since the non-linear backreaction within the non-Abelian gauge sector tends to inhibit the growth of the non-Abelian gauge fields.
In this case, it may be possible to accurately describe the dynamics of the inflaton and Abelian gauge sector, despite the loss of perturbativity in the non-Abelian sector. We leave a more detailed investigation of this setup for future work.

Let us finally comment on the possible impact of charged particles.
Our discussion so far has neglected such contributions, implicitly assuming that the gauge groups belong to a dark sector with no or only very heavy particles charged under these gauge symmetries.
 A more attractive scenario however may be attempt to identify these gauge groups with SM gauge groups, in which case we need to consider the effect of many charged massless degrees of freedom (assuming the electroweak symmetry to be unbroken). In the presence of a chiral anomaly (as in the SM) this leads to a dual production of gauge fields and massless fermions during inflation~\cite{Domcke:2018eki,Domcke:2018gfr}. For Abelian gauge theories, this leads to an induced fermion current which strongly backreacts on the gauge field production, significantly inhibiting the latter~\cite{Domcke:2018eki}. On the contrary, in non-Abelian gauge theories this process does not significantly impact the predictions for the gauge field production~\cite{Domcke:2018gfr}, mainly because the backreaction of the induced fermion current is subdominant compared to the backreaction of the non-trivial gauge field background. For the production of massive fermions in the context of Abelian and non-Abelian axion inflation see e.g.\ Refs.~\cite{Adshead:2015kza,Adshead:2018oaa,Min:2018rxw,Mirzagholi:2019jeb}.


\section{Conclusion}
\label{sec:conclusion}

The early Universe inflationary paradigm may be explained naturally in the sense of t'Hooft with an axion acting as the inflaton.
This allows in particular a Chern-Simons coupling to non-Abelian gauge fields.
Such an interaction is not only the unique dimension $5$ coupling of the axion to gauge fields which respects the axion's approximate shift symmetry, but it also yields phenomenological rich implications such as the production of maximally chiral gravitational waves (at linear level in cosmic perturbation theory).

We show in a broad model parameter space under which conditions gauge field fluctuations can source a transition to a stable non-zero gauge field background.
In particular this applies also if the gauge field dynamically evolve from the Bunch-Davies vacuum in the far past, i.e.\ without invoking any additional mechanism generating the gauge field background.
As the main point of this paper we derive constraints which the model has to face due to (i) theoretical consistency and (ii) numerical restrictions.
Let us make this more explicit in the following.

We require for theoretical consistency (a) the fundamental axion decay constant to be sub-Planckian $f_{a} < M_{P}$ and (b) the gauge group coupling to be perturbative $g \leq 1$.
To realize (a) together with super-Planckian axion field excursion required by the Planck data (for monomial potentials), we investigate two representative options:
The alignment of multiple Peccei-Quinn axions {as in N-flation models}~\cite{Dimopoulos:2005ac,Kim:2004rp} and the (un)winding of moduli space in string theoretical axions as in monodromy~\cite{McAllister:2008hb,Silverstein:2008sg}.
We demonstrate that both options are strongly constrained either due to the appearance of a light degree of freedom which spoils the EFT or due to the axion field excursion requiring excessively large winding numbers,
$\Delta a/f_{\axion} \gg 10^3$.
The remaining theoretically consistent parameter space which allows for the formation of the non-trivial gauge field background is in the case of Peccei-Quinn axion models restricted to $g \gtrsim 5 \times 10^{-2}$ with $f_\axion \gtrsim 10^{-5}~M_P$ and for axion monodromy to $g \gtrsim 0.5$ with $f_\axion \gtrsim 10^{-3}~M_P$.
For small $g$ the constraints enforce a decoupling between axion and gauge field sector.
Hence natural inflation is recovered in this regime.
On the other hand we find in the regime of large $g$ and small $g^2/f_a$ that the non-Abelian model remains in its Abelian limit essentially over the whole course of inflation.

For the numerical analysis we are restricted to a linearized treatment of the gauge field perturbations. 
This in particular limits our ability to make quantitative predictions in the regime where the two (effective) couplings, $g$ and $g^2/f_a$ are large.
 We hope that our work will trigger further investigations of this part of the parameter space, possibly by means of dedicated lattice simulations. 
 This requires to overcome the difficulty of simulating this system in exponentially expanding de Sitter spacetime, where lattice simulations can currently only be obtained for a few e-folds~\cite{Cuissa:2018oiw}, insufficient for our purposes.

As an outlook, we consider simultaneous axion interactions with both Abelian and non-Abelian gauge fields, as may be expected in the SM. For similar values of the two gauge couplings, the backreaction onto the inflaton sector is dominated by the Abelian gauge fields, thus circumventing some of the obstacles of a full non-linear analysis. This conclusion may however be altered if light fermions are included in the model, since they significantly inhibit only the growth of the Abelian gauge fields~\cite{Domcke:2018eki,Domcke:2018gfr}.


\paragraph{Acknowledgements}
We thank Azadeh Maleknejad and Alexander Westphal for fruitful discussions and helpful comments on the manuscript. 
Also we thank Mauro Pieroni and Francesco Muia for carefully reading the manuscript and for the helpful comments.
This work was funded by the Deutsche Forschungsgemeinschaft (DFG) under Germany's Excellence Strategy - EXC 2121 Quantum Universe - 390833306.


\newpage

\appendix

\section{Approximations and Uncertainties}
\label{app:approximation}

In the linearized analysis for the $SU(2)$ gauge group we expand the gauge fields around an isotropic and homogeneous background as~\cite{Maleknejad:2011jw,Maleknejad:2011sq,Adshead:2012kp,Domcke:2018rvv}
\begin{align}
    A^{b}_{i}(t,\mathbf{x}) &= R(t)\psi(t)\delta_{i}^{b} + \delta A_{i}^{b}(t,\mathbf{x}).
\end{align}
For constant $\xi \geq 2$ the gauge fields may develop a non-zero vev {at late times} which can be analytically approximated by
\begin{align}
\psi = H\frac{c_{i}\xi}{g},
\end{align}
with $c_{i} = \{c_{1},c_{2} \}$ as given in Eq.~(\ref{eq:csolutions}).
We show in this appendix
that this expression with $c_{i} = c_{2}$ very well approximates the numerical solution by directly solving the coupled system of {homogeneous} Eqs.~(\ref{eq:su2fulleom}) and (\ref{eq:eomgamma}).

Therefore, let us first comment on the matching procedure we followed in the main text.
Recall, that we derived that the Abelian fluctuations will trigger the transition to a stable non-zero vev when the threshold
\begin{align}
    \label{appeq:transition}
    -\tau g\sqrt{\langle A_{\text{AB}}^2} \rangle \sim c_{2} \xi
\end{align}
is reached, see Eq.~(\ref{eq:transition}).
This however, may be too conservative since also smaller fluctuations may reach the $c_{1}$ solution (which can be interpreted as a saddle point as it forms a one parameter family only) and then eventually evolve into the $c_{2}$ solution. 
The resulting uncertainty in the exact matching time increases with decreasing gauge group coupling $g$, 
 such that the largest uncertainty falls into the magnetic drift regime.
Thus, we have to treat this regime with special care in the numerical evaluation (although we find this regime to be theoretically inconsistent in the UV completions considered in the main text).
When matching too late, the background field enters a strongly oscillating phase 
over a large time period in the numerical evaluation, c.f. Fig.~\ref{fig:flowlines}.
This in turn has a strong non-physical impact in the axion evolution as it simulates a 
non-monotonic axion evolution -- but we should strictly have $a' < 0$ (in our convention).
So, we conclude that for the magnetic drift regime it is necessary to change to the inherently non-Abelian description earlier than indicated by Eq.~\eqref{eq:transition}.
A natural replacement in Eq.~(\ref{appeq:transition}) is thus given by $c_{2} \mapsto c_{1}$ .
Note again, that this treatment is only needed {(deep) inside } the magnetic drift regime. 
Outside of this regime
the difference in the matching time is negligible and thus our procedure used so far is well justified, c.f. Fig.~\ref{appfig:outsidebackground}.
\begin{figure}[!h]
  \centering
    \subfigure{\includegraphics[width = 7cm]{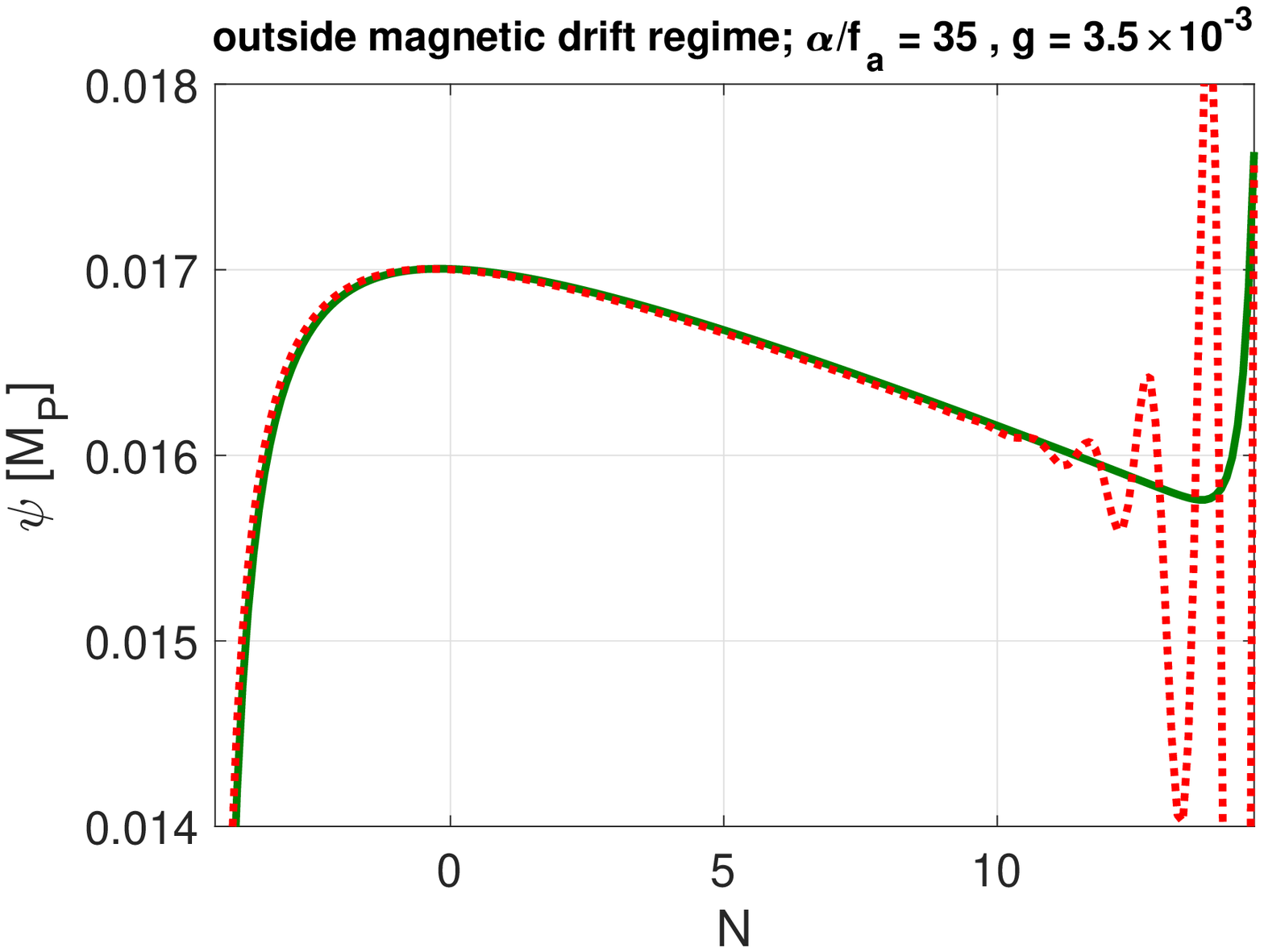}}
    \subfigure{\includegraphics[width = 7cm]{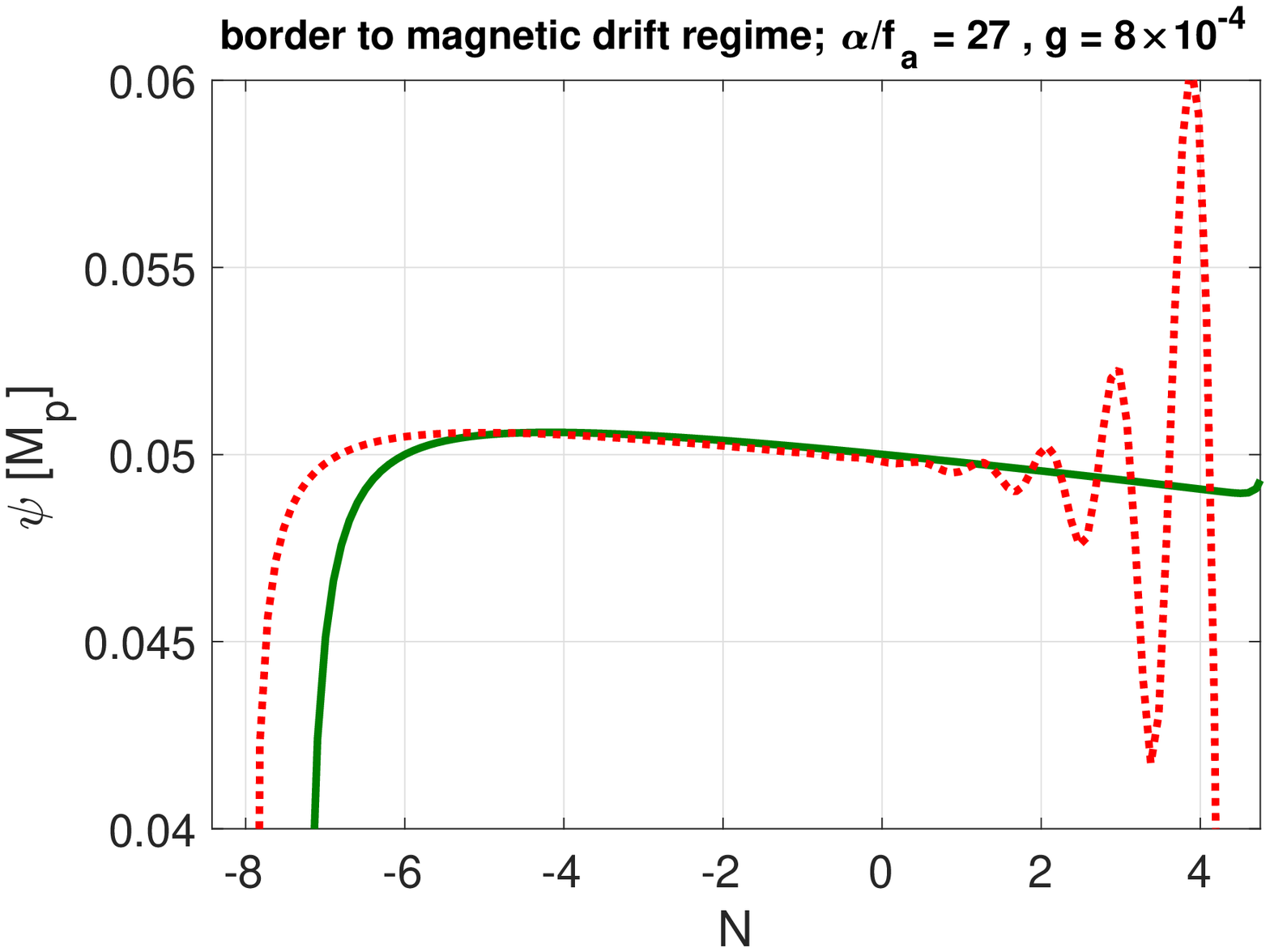}}
    \caption{Analytic background field with the $c_{2}$ solution (solid green) compared to its numerical solution (dashed red). The matching condition \eqref{appeq:transition} is sufficient for large gauge group couplings and/or small effective interaction couplings. 
     }
    \label{appfig:outsidebackground}
\end{figure}
This figure also clearly demonstrates that the (small amplitude) oscillatory regime is left within $ \sim 2$ e-folds after the matching.
Let us now turn to the background field (deep) inside the magnetic drift regime.
We demonstrate in Fig.~\ref{appfig:mdrbackground} that matching the Abelian fluctuations to the $c_2$ solution leads to a too late matching point
-- causing an oscillation with a high amplitude.
The problem is resolved by an earlier matching to
the $c_{1}$ solution as depicted in the same figure.
\begin{figure}[!h]
  \centering
    \subfigure{\includegraphics[width = 7cm]{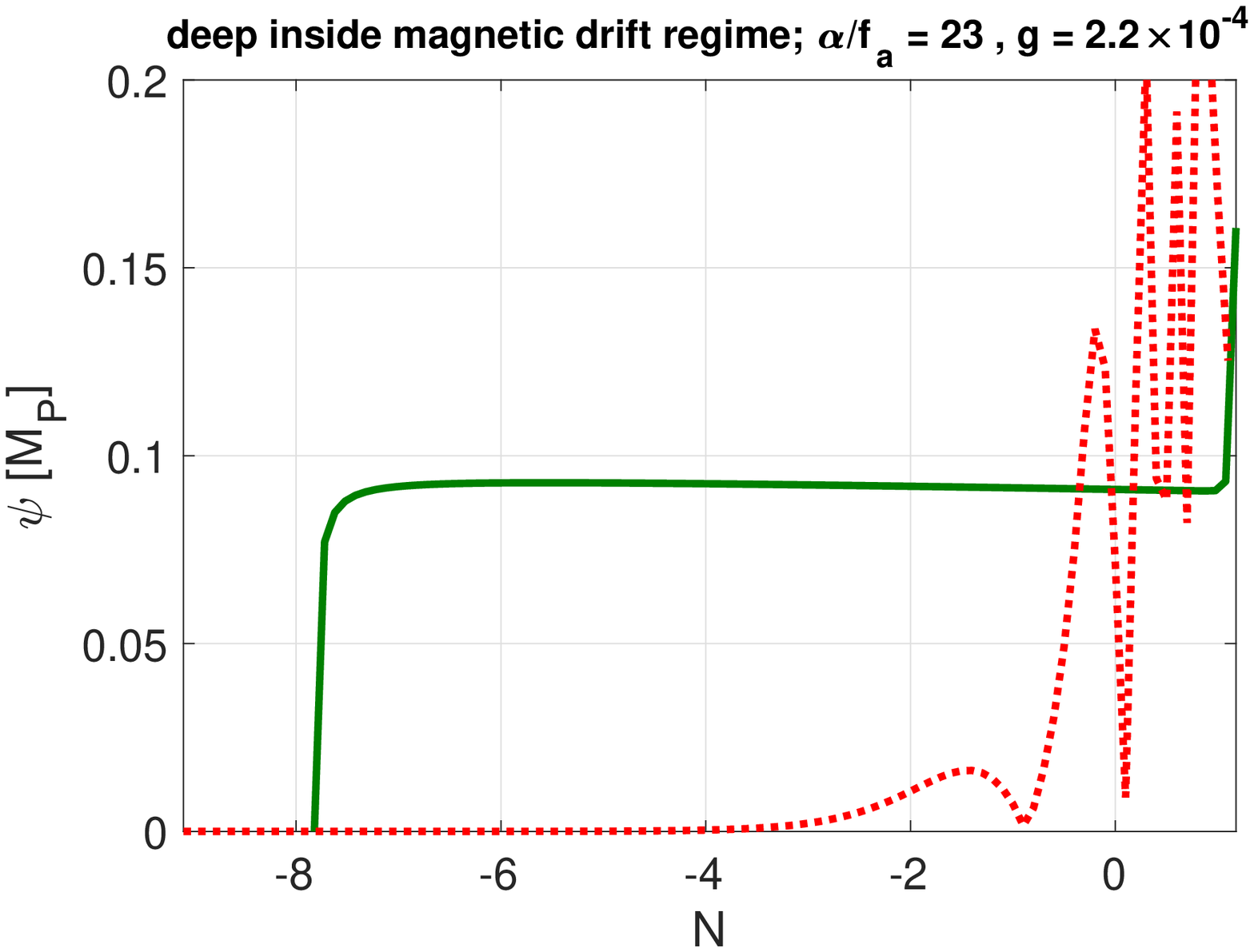}}
    \subfigure{\includegraphics[width = 7cm]{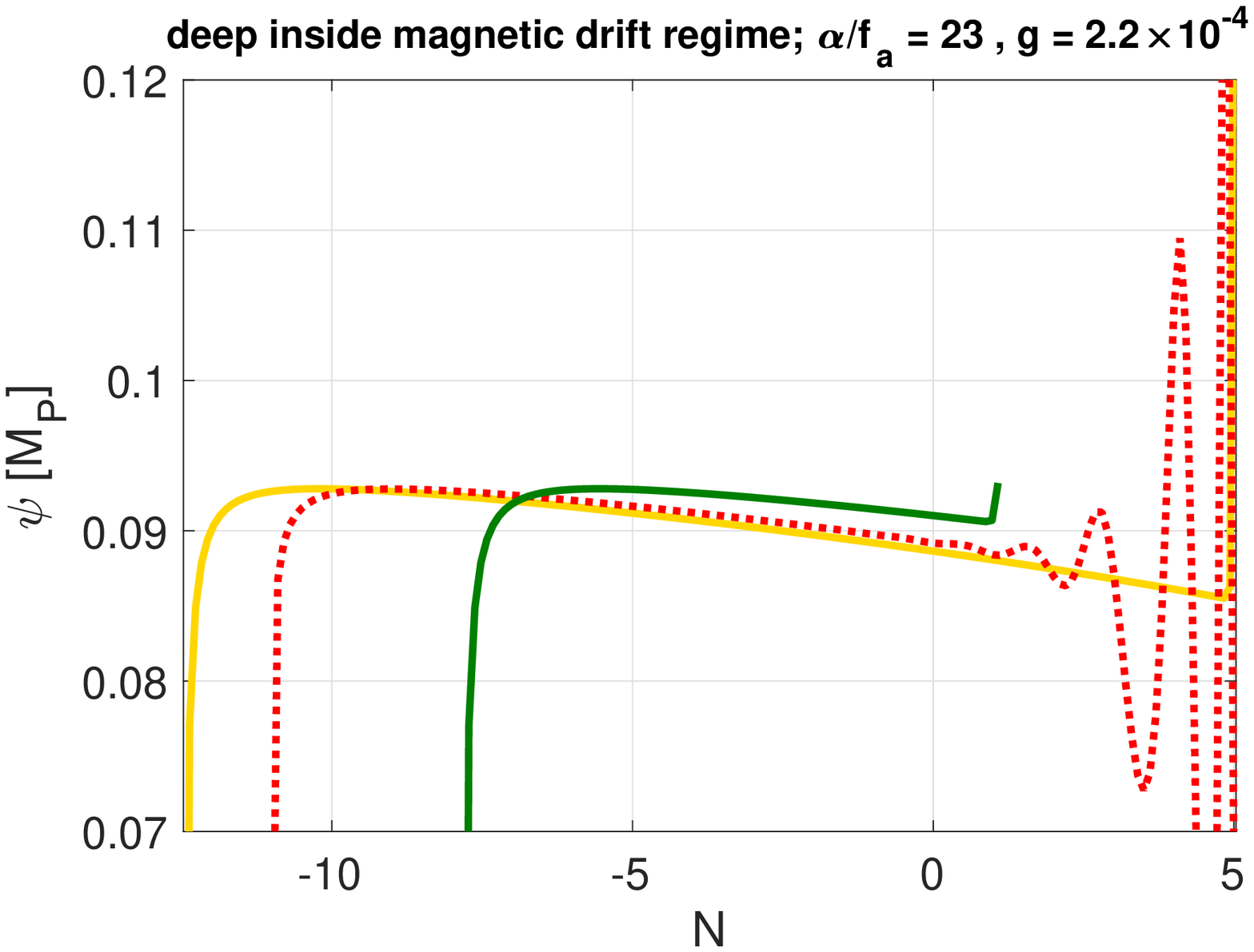}}
    \caption{\textbf{Left}: Matching to the $c_{2}$ solution deep inside the magnetic drift regime is accompanied with a strong oscillatory phase caused by the too late matching. 
    This causes a (unphysical) non-monotonic evolution of the axion and the (numerically evaluated) gauge field background jumps to the zero solution (red dashed).
    \textbf{Right}: An earlier matching with the $c_{1}$ solution damps the oscillation amplitude in the numerical gauge field background evaluation (red dashed). 
    We compare to the $c_{2}$ solution analytic gauge field background which we get when matching with the $c_{1}$ solution (solid yellow) and the standard criterion with the $c_{2}$ solution (solid green in both plots).
     }
    \label{appfig:mdrbackground}
\end{figure}
Note that in both plots the convention for the e-fold $N$ is different than the one in the main text.
Here we do not define $N=0$ as the time when $\epsilon = 1$, as we are not interested in the CMB scales (for which the shift in the e-fold may be important).
Rather, $N=0$ is set by $\epsilon_{\text{vac}} = 1$. 
That it why we have $N<0$ as the end of inflation in these plots (caused by the gauge field induced friction).


\newpage
\bibliographystyle{JHEP}
\bibliography{refs}{}

\end{document}